\documentclass[useAMS,usenatbib,referee]{biom}
\usepackage[utf8]{inputenc}
\usepackage{amsmath}
\usepackage{booktabs}
\usepackage{multirow}
\usepackage{amsfonts}
\usepackage{amssymb}
\usepackage{tabularx}
\usepackage{graphicx}
\usepackage{placeins}
\usepackage{afterpage}
\usepackage{setspace}
\usepackage{mathtools}
\usepackage{pdfpages}
\usepackage{listings}
\usepackage{bm,bbm}
\usepackage{mathrsfs}
\usepackage{url}
\usepackage{mdframed}
\usepackage[normalem]{ulem}
\usepackage{xcolor}
\usepackage{soul}
\usepackage{array} 
\usepackage{romannum}
\usepackage{verbatim}
\usepackage{siunitx}
\usepackage[makeroom]{cancel}
\usepackage{xr}
\setstcolor{red}

\usepackage{titlesec}
\titleformat{\section}
  {\normalfont\Large\bfseries}{\thesection}{1em}{}
\titlespacing*{\section}
  {0pt}{8pt}{0pt} 
\titlespacing*{\subsection}
  {0pt}{0pt}{0pt} 
\titlespacing*{\subsubsection}
  {0pt}{0pt}{0pt} 
  
\AtBeginDocument{\pagenumbering{arabic}}

\newcommand{\blind}{0}

\if0\blind
{
  \title[A Likelihood Approach to Incorporating Self-Report Data in HIV Recency Classification]{A Likelihood Approach to Incorporating Self-Report Data in HIV Recency Classification}  
  \author{Wenlong Yang$^{1,*}$\email{wenlong@psu.edu},
  Danping Liu$^{2,**}$\email{danping.liu@nih.gov},
  Le Bao$^{1,***}$\email{lub14@psu.edu},
  Runze Li$^{1,****}$\email{rzli@psu.edu} \\
$^{1}$Department of Statistics, Pennsylvania State University at University Park, Pennsylvania, U.S.A. \\
$^{2}$Biostatistics Branch, Division of Cancer Epidemiology and Genetics, National Cancer Institute, U.S.A.}
}
\fi

\if1\blind
{
  \bigskip
  \bigskip
  \bigskip
  \begin{center}
    {\LARGE\bf A Likelihood Approach to Incorporating Self-Report Data in HIV Recency Classification}
\end{center}
  \medskip
} \fi

\begin{document}
\newcommand\redsout{\bgroup\markoverwith{\textcolor{red}{\rule[0.5ex]{2pt}{0.4pt}}}\ULon}

\begin{abstract}
Estimating new HIV infections is significant yet challenging due to the difficulty in distinguishing between recent and long-term infections. We demonstrate that HIV recency status (recent v.s. long-term) could be determined from self-report testing history and biomarkers, which are increasingly available in bio-behavioral surveys.
HIV recency status is partially observed, given the self-report testing history. For example, people who tested positive for HIV over one year ago should have a long-term infection. Based on the nationally representative samples collected by the Population-based HIV Impact Assessment (PHIA) Project, we propose a likelihood-based probabilistic model for HIV recency classification. The model incorporates individuals with known recency status based on testing histories
and individuals whose recency status could not be determined and integrates the mechanism of how HIV recency status depends on biomarkers and the mechanism of how HIV recency status, together with the self-report time of the most recent HIV test, impacts the test results. We compare our method to logistic regression and the binary classification tree (current practice) on Malawi PHIA data, as well as on simulated data. Our model obtains more efficient and less biased parameter estimates and is relatively robust to potential reporting error and model misspecification. 
\end{abstract}


\maketitle


\section{Introduction}
\label{sec:Intro}

HIV remains a global health crisis. In 2021, HIV claimed 650,000 lives worldwide and was carried by 38.4 million people at the end of the year, highlighting the ongoing struggle to control this pandemic pathogen. Compared with the notable progress in reducing AIDS-related deaths via antiretroviral therapy (ART), efforts to prevent new infections have been less successful
. Global HIV incidence per 1000 population among adults ($15+$) remains at 0.23 in 2021 and has only declined by 39\% since 2010, missing the UN General Assembly's target of 75\% reduction for 2020. Despite great efforts, progress against HIV remains inadequate and fragile in many countries, such as those in sub-Saharan Africa that account for 57\% of new HIV infections in 2021 \citep{aidsinfo2023}.

HIV incidence, or new HIV infections, refers to the estimated proportion of people with a newly acquired HIV infection within a susceptible population. Prospective cohort studies can directly monitor incident HIV infections by longitudinal follow-up of susceptible population, but are costly and prone to selection bias, changes in behavior associated with study participation, and nonrandom dropout \citep{laeyendecker2013hiv,fellows2020}. 
An alternative way to estimate HIV incidence is through cross-sectional surveys, but this requires an algorithm to classify HIV-positive individuals into recent cases (within the past year) and long-term cases. Specifically, a recent infection testing algorithm (RITA) classifies each survey participant's HIV recency status (recent versus long-term infection) based on biomarkers \citep{world2022using}. The proportion of individuals classified as recent infection needs to be adjusted for two parameters: mean duration of recent infection (MDRI), which is the average time an individual with a recent infection remains classified as a recent case by the RITA and is ideally within the range of 4–12 months, and the false recency rate (FRR) of the RITA \citep{kassanjee2012new}. Both MDRI and FRR are known to be highly context-dependent, and inappropriate parameter values might lead to questionable HIV incidence estimate \citep{Fellows2023jv}.

Since 2014, the Population-based HIV Impact Assessment (PHIA) Project has conducted HIV-focused, cross-sectional, household-based, nationally representative surveys to measure the reach and impact of HIV programs in countries seriously affected by HIV. These cross-sectional surveys use a stratified multistage survey sampling design. Data collection consists of household interviews, individual interviews, and laboratory testing. Interview questions cover household membership and characteristics, individual sociodemographic characteristics, and HIV-related risk factors. Voluntary biomarker testing is offered to whoever completes an individual interview. Blood samples from individuals who test positive for HIV undergo additional testing of HIV RNA viral load, CD4+ cell count, antiretroviral (ARV) drug presence, etc. More details of PHIA datasets will be presented in Section~\ref{sec:Data}.

The current practice of HIV incidence estimation in the PHIA Project is the RITA that classifies an HIV-positive individual as 
a recent case (within one year) if the normalized optical density is low (ODn $\leq 1.5$) in HIV-1 limiting antigen avidity enzyme immunoassay, the viral load is high (VL $\geq 1000$ copies/mL), and no ARV has been detected \citep{icap2021phia}. It sometimes classifies all individuals as long-term cases because recent infections are rare events among HIV-positive individuals, and the sample size of HIV-positive individuals is small, especially when dis-aggregating the data by age groups or sub-national areas.
\cite{Sheng2023} was the first probabilistic classification model for analyzing the recent infections in PHIA surveys. Their idea was to borrow information from external cohort studies where the relationship between HIV recency status and the covariates are presented in a contingency table. However, this approach relies on the strong and untestable assumption that the external data sets and PHIA data exhibit the same relationship between recency status and biomarkers.

In this paper, we proposed a likelihood-based inference to estimate the probability of recent HIV infections, by utilizing self-reported HIV test history. We argue that self-report data provide important information beyond the biomarkers used by RITA or \cite{Sheng2023}, which should not be overlooked. Specifically, a self-reported positive HIV test over one year ago would indicate a long-term infection; a self-reported negative HIV test within one year would translate to a recent infection; HIV recency status of other self-reporting patterns can not be inferred directly. Hence, HIV recency status can be inferred only for those with
certain combinations of whether tested for HIV within one year and the test result (positive v.s. negative).
A likelihood-based probabilistic model is then used to integrate the mechanism of how the result of the last HIV test depends on the testing time and HIV recency status and the mechanism of how HIV recency status depends on biomarkers. 
For the former mechanism, we specify the conditional distribution of the test result based on the fact that a person with HIV is more likely to have received positive test results if the test took place later in time. For the latter mechanism, we assume a logistic regression model of recency status given biomarkers. An important advantage of our proposed method is that it does not require the use of external data sets or the specification of global parameters of MDRI and FRR.

The rest of the paper is organized as follows. In Section~\ref{sec:Data}, we briefly introduce the PHIA study design and the analytic dataset. Section~\ref{sec:Method} proposes the likelihood inference and 
\textcolor{black}{examines} the underlying key model assumptions. Then in Section~\ref{sec:realdata}, we present the results from both our model and current practice on Malawi PHIA data. In Section~\ref{sec:sim}, we compare our model to two competing statistical models using a series of simulation studies, and examine our model performance at the presence of reporting error and model misspecification. Finally, we make concluding remarks in Section~\ref{sec:discuss}.

\section{Data Source and Preprocessing}
\label{sec:Data}
In this section, we introduce the Malawi PHIA data \citep{mphia2016} and how we choose eligible individuals for the study. Then we briefly explain the data processing procedures before modeling.

Malawi PHIA surveys were conducted from 2015 to 2016. PHIA used a complex survey design, stratified on sub-national geographic divisions, such as regions or provinces, as identified in the most recent census of each country. Within each stratum, census enumeration areas (EAs) are randomly selected with probability proportional to population size (1st stage). Subsequently, a random sample of households is selected within selected EAs (2nd stage). Consented households were administered a household interview completed by the household head. Within a selected household, all eligible adults aged 15 years or older were given an individual interview and voluntary biomarker testing. After data collection, PHIA calculated a sampling weight for each individual participating in the blood test, accounting for sampling probabilities, nonresponse, and undercoverage of certain population groups.

We consider the PHIA covariates that are the most relevant to HIV recency classification, including age, gender, normalized optical density (ODn), viral load (VL), and CD4 count. The biomarkers were available for those who completed the individual interview and consented to undergo blood draw. In the individual interview, participants were asked about their most recent HIV testing history, which included whether they had taken HIV tests and received definite results in the past year, the year and month of their last test, and the associated test results (positive or negative). Note that for privacy considerations, PHIA data redact the day of self-reported HIV tests and the day of individual interviews.

As recommended by PHIA reports, individuals on ART are believed to have long-term (over one year ago) infections \citep{voetsch2021hiv}, though this assumption is becoming less reliable in recent years. Therefore, our interest is to estimate the HIV recency status for the population that are living with HIV but not on ART. 

Our analytic sample included subjects who were: (1) at least 15 years old, (2) HIV-positive at the time of the survey, (3) not on ART based on self-report or lab-detection, (4) having positive sampling weights, and (5) having self-reported HIV testing history and biomarkers. This resulted in $n=408$ subjects in the analytic sample (see Web Figure 1 for a CONSORT diagram describing sample selection). We then rescale the individual sampling weights so that they sum up to 408. 
For less than 8\% of the individuals, VL is categorical. Following \cite{Sheng2023}, we set ``undetectable" to 20 and ``less than 40" to 30. \textcolor{black}{As suggested in \cite{stirrup2019associations}, we work with the logarithm of VL (denoted $\text{logVL}$) and the square root of CD4 count (denoted CD4)}. All the continuous variables (age, ODn, logVL, and CD4) were also standardized so that they had mean 0 and standard deviation 1.

\section{Proposed Model}
\label{sec:Method}

This section introduces the likelihood construction for HIV recency status and HIV testing history. We use subscript $i=1,\cdots,n$ to index individuals. When there is no confusion, the subscript is omitted for simplified notation. Use $Z$ to denote the last test results (1 for positive and 0 for negative), and $S$ to denote the time gap measured by year between the last test and PHIA interview. Let $\boldsymbol{X}$ denote individual covariates including 1 (for intercept), age, gender \textcolor{black}{(1 for male and 0 for female)}, ODn, logVL, and CD4, and $Y$ denote the true HIV recency status (1 for recent infection within one year and 0 for long-term infection over one year ago). When the last HIV test was in the same month as the interview, we set $S=10/365$, which is the expected value of the time gap if we assume the two days are uniformly distributed over the triangle with the vertices at $(0,0)$, $(0,30)$, $(30,30)$.

We are interested in estimating how HIV-positive individuals' recency status relates to age, gender, and biomarkers ($E(Y|\boldsymbol{X})$), and the population-average proportion of recent infections ($E(Y)$). 
In some cases, HIV test time and results together could determine HIV recency status: a person who tested positive for HIV one year ago must have acquired HIV beyond one year, i.e., $S>1$ and $Z=1$ implies that $Y=0$; simiarly, $S\leq1$ and $Z=0$ implies that $Y=1$. When HIV recency status can not be inferred from test time and results, those are still informative. For example, a subject who was tested positive 11 months ago is very likely to have a long-term infection. Figure~\ref{fig1} shows that 46\% of included PHIA individuals are associated with a definite HIV recency status.
\begin{figure}[!h] 
    \centerline{\includegraphics[width=0.5\textwidth]{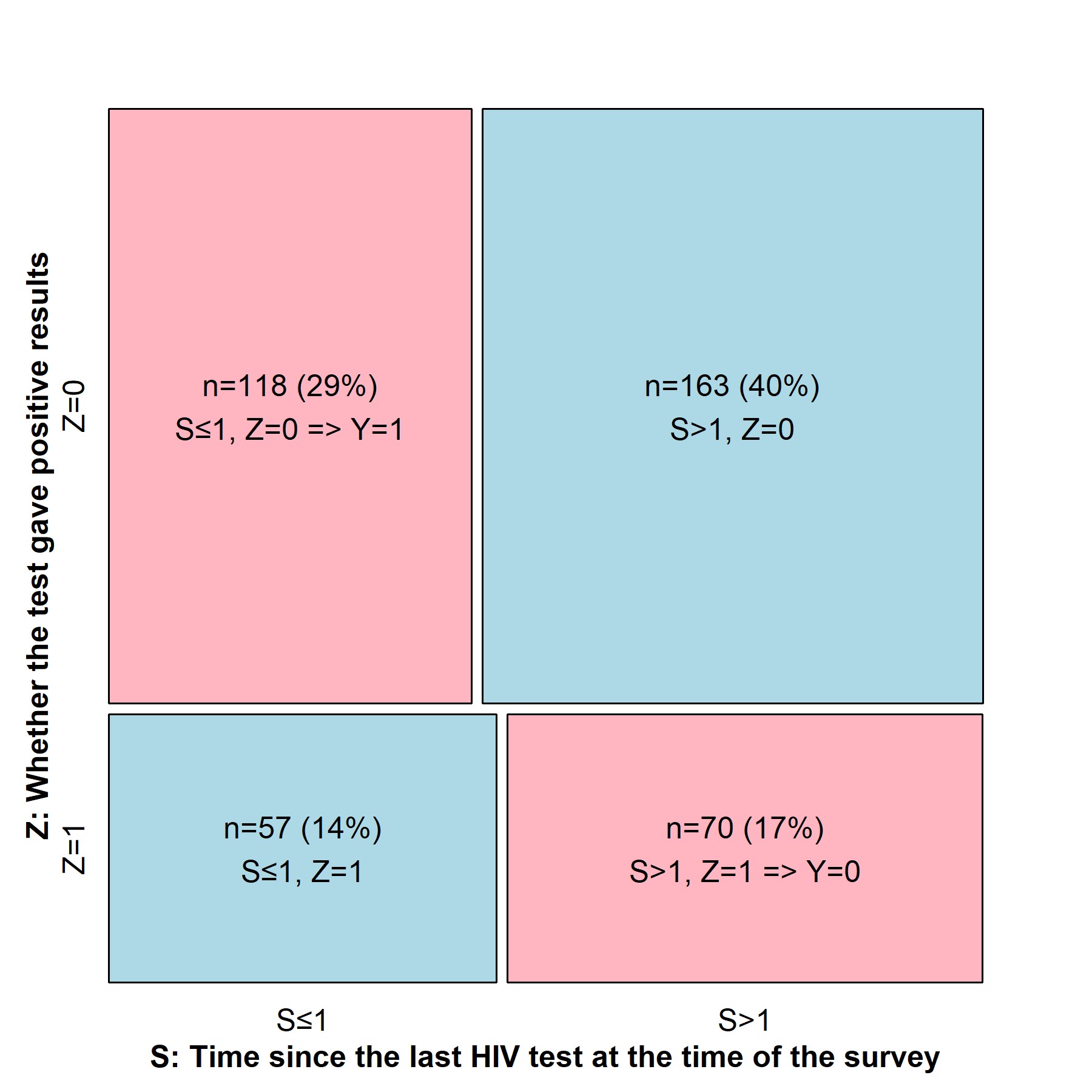}}
    \caption{Mosaic plot showing whether the subject received a test within the past year and whether the test result was positive, based on HIV-positive participants not on ART in the Malawi PHIA dataset. Top left and bottom right blocks represent data with observed responses, while top right and bottom left ones represent unobserved responses.}
    \label{fig1}
\end{figure}

Section 3.1 shows the likelihood construction when we assume a parametric model for $S\mid Y,\boldsymbol{X}$. We then introduce a semiparametric alternative model for $S\mid Y,\boldsymbol{X}$ in Section 3.2. Section 3.3 presents two types of risk to predict the probability of recent HIV infection.

\baselineskip=25pt

\subsection{Parametric Likelihood Inference}
\label{subsec:lik}

The joint conditional distribution of $Z$, $S$, and $Y$ given $\boldsymbol{X}$ can be written as:
\[
f_{Z,S,Y\mid \boldsymbol{X}}(z,s,y\mid \boldsymbol{x})=P(Y=y\mid \boldsymbol{X}=\boldsymbol{x})f_{S\mid Y,\boldsymbol{X}}(s\mid y,\boldsymbol{x})P(Z=z\mid S=s,Y=y,\boldsymbol{X}=\boldsymbol{x}).
\]

For $P(Y=y\mid \boldsymbol{X}=\boldsymbol{x})$, we assume a logistic regression model:
\begin{equation}
\pi_{i}:= P\left(Y_{i}=1 \mid \boldsymbol{X}_{i}=\boldsymbol{x}_{i}\right)={\exp \left(\boldsymbol{x}_{i}^{\top} \boldsymbol{\beta}\right)}/{\{1+\exp \left(\boldsymbol{x}_{i}^{\top} \boldsymbol{\beta}\right)\}},
\label{eq:pi_i}
\end{equation} where $\boldsymbol{\beta}$ are unknown parameters. This model is of our primary interest, which estimates the covariates associated with recent infection.

For $f_{S\mid Y,\boldsymbol{X}}(s\mid y,\boldsymbol{x})$, we assume a Gamma distribution with the shape parameter fixed and the scale parameter depending on $Y,\boldsymbol{X}$:
\begin{equation}
    S\mid Y,\boldsymbol{X}\sim\operatorname{Gamma}(\alpha,\lambda_{Y,\boldsymbol{X}}),\quad
    \log(\lambda_{Y,\boldsymbol{X}})=\xi_YY+\boldsymbol{\xi}^\top\boldsymbol{X}.
    \label{eq:gamma}
\end{equation}

For $P(Z=z\mid S=s,Y=y,\boldsymbol{X}=\boldsymbol{x})$, certain combinations of $S$ and $Y$ lead to definite test results with probability 1: $P(Z_i=1\mid S_i\leq 1,Y_i=0)=1$, $P(Z_i=0\mid S_i> 1,Y_i=1)=1$.
That is, for people with long-term HIV infections ($Y=0$), if their previous HIV tests were taken within one year ($S\leq 1$), then the result must be positive ($Z=1$); for people with recent HIV infections, if their previous HIV tests were taken more than one year ago, then the result must be negative. For the remaining people, we assume that the earlier the test took place, the less likely it was to give a positive result with a log link model:
\begin{equation}
    \begin{aligned}
      p_{0 i} &:= P\left(Z_{i}=1 \mid S_i=s_i>1,Y_{i}=0, \boldsymbol{X}_{i}=\boldsymbol{x}_{i}\right) = \operatorname{exp}(\eta_{0}\log s_{i}),\\
      p_{1 i} &:= P\left(Z_{i}=1 \mid S_i=s_i\leq 1,Y_{i}=1, \boldsymbol{X}_{i}=\boldsymbol{x}_{i}\right)=1-\operatorname{exp}(\eta_{1}\log s_{i}),\\
    \end{aligned}
    \label{eq:p0i_p1i}
\end{equation}
where $S$ is considered on a log scale because it is right skewed, $\eta_{0}<0$ and $\eta_{1}>0$ are nuisance parameters ensuring $p_{0 i}$ and $p_{1 i}$ are both decreasing functions of $S_i$. A discussion of other choices of link functions in Equation~(\ref{eq:p0i_p1i}) is included in Web Appendix A. Here we could also include $\boldsymbol{X}_{i}$ as a predictor of $Z_i$. However, in our application, it is unlikely that the self-report test result $Z_i$ was affected by age, gender, ODn, logVL, or CD4, so we only included $S_{i}$.  

Upon observing $S$ and $Z$, we can show each subject's likelihood contribution:

\textbf{\Romannum{1}. Subjects with the latest test within one year and negative test result ($S\leq 1,Z=0\implies Y=1$)}
\begin{equation*}
\begin{aligned}
    f_{Z,S,Y\mid \boldsymbol{X}}(z_i,s_i,y_i\mid \boldsymbol{x}_i)    =&P(Y_i=1\mid\boldsymbol{X}_i=\boldsymbol{x}_i)f_{S\mid Y,\boldsymbol{X}}(s_i\mid 1,\boldsymbol{x}_i)P(Z_i=0\mid S_i=s_i,Y_i=1,\boldsymbol{X}_i=\boldsymbol{x}_i)\\
    =&\pi_if_{S\mid Y,\boldsymbol{X}}(s_i\mid 1,\boldsymbol{x}_i)(1-p_{1i}).
\end{aligned}
\end{equation*}

\textbf{\Romannum{2}. Subjects with the latest test over one year ago and positive test result ($S>1,Z=1\implies Y=0$)}
\begin{equation*}
\begin{aligned}
    f_{Z,S,Y\mid \boldsymbol{X}}(z_i,s_i,y_i\mid \boldsymbol{x}_i)=&P(Y_i=0\mid\boldsymbol{X}_i=\boldsymbol{x}_i)f_{S\mid Y,\boldsymbol{X}}(s_i\mid 0,\boldsymbol{x}_i)P(Z_i=1\mid S_i=s_i,Y_i=0,\boldsymbol{X}_i=\boldsymbol{x}_i)\\
    =&(1-\pi_i)f_{S\mid Y,\boldsymbol{X}}(s_i\mid 0,\boldsymbol{x}_i)p_{0i}.
\end{aligned}
\end{equation*}

\textbf{\Romannum{3}. Subjects with the latest test within one year and positive test result ($S\leq 1,Z=1,Y$ unknown)} 
\begin{equation*}
\begin{aligned}
    f_{Z,S,Y\mid \boldsymbol{X}}(z_i,s_i,y_i\mid \boldsymbol{x}_i)=&P(Y_i=y_i\mid\boldsymbol{X}_i=\boldsymbol{x}_i)f_{S\mid Y,\boldsymbol{X}}(s_i\mid y_i,\boldsymbol{x}_i)P(Z_i=1\mid S_i=s_i,Y_i=y_i,\boldsymbol{X}_i=\boldsymbol{x}_i)\\
    =&\begin{cases}
    (1-\pi_i)f_{S\mid Y,\boldsymbol{X}}(s_i\mid 0,\boldsymbol{x}_i) & \text{if }y_i=0\\
    \pi_if_{S\mid Y,\boldsymbol{X}}(s_i\mid 1,\boldsymbol{x}_i)p_{1i} & \text{if }y_i=1    
    \end{cases}.
\end{aligned}
\end{equation*}
The likelihood contribution is thus $(1-\pi_i)f_{S\mid Y,\boldsymbol{X}}(s_i\mid 0,\boldsymbol{x}_i)+\pi_if_{S\mid Y,\boldsymbol{X}}(s_i\mid 1,\boldsymbol{x}_i)p_{1i}$.

\textbf{\Romannum{4}. Subjects with the latest test over one year ago and negative test result ($S>1,Z=0,Y$ unknown)}
\begin{equation*}
\begin{aligned}
    f_{Z,S,Y\mid \boldsymbol{X}}(z_i,s_i,y_i\mid \boldsymbol{x}_i)    =&P(Y_i=y_i\mid\boldsymbol{X}_i=\boldsymbol{x}_i)f_{S\mid Y,\boldsymbol{X}}(s_i\mid y_i,\boldsymbol{x}_i)P(Z_i=0\mid S_i=s_i,Y_i=y_i,\boldsymbol{X}_i=\boldsymbol{x}_i)\\
    =&\begin{cases}
    (1-\pi_i)f_{S\mid Y,\boldsymbol{X}}(s_i\mid 0,\boldsymbol{x}_i)(1-p_{0i}) & \text{if }y_i=0\\
    \pi_if_{S\mid Y,\boldsymbol{X}}(s_i\mid 1,\boldsymbol{x}_i) & \text{if }y_i=1   
    \end{cases}.
\end{aligned}
\end{equation*}
The likelihood contribution is thus $(1-\pi_i)f_{S\mid Y,\boldsymbol{X}}(s_i\mid 0,\boldsymbol{x}_i)(1-p_{0i})+\pi_if_{S\mid Y,\boldsymbol{X}}(s_i\mid 1,\boldsymbol{x}_i)$.

To conclude, we derive the observed data likelihood as (note that in PHIA data, each subject $i$ is associated with a sampling weight $w_i$):
\begin{equation}
\begin{aligned}
\mathcal{L}(\boldsymbol\theta)&=\prod_{i=1}^n f_{Z,S\mid \boldsymbol{X}}(z_i,s_i\mid \boldsymbol{x}_i)^{w_i}=\prod_{i=1}^n\left\{\sum_{j=0}^1 f_{Z,S,Y\mid \boldsymbol{X}}(z_i,s_i,j\mid \boldsymbol{x}_i)\right\}^{w_i}\\
&=\prod_{i=1}^n \left\{\pi_if_{S\mid Y,\boldsymbol{X}}(s_i\mid 1,\boldsymbol{x}_i)(1-p_{1i})\right\}^{\mathbbm{1}(s_i\leq 1)(1-z_i)w_i}\left\{(1-\pi_i)f_{S\mid Y,\boldsymbol{X}}(s_i\mid 0,\boldsymbol{x}_i)p_{0i}\right\}^{\mathbbm{1}(s_i>1)z_iw_i}\times\\
&\left\{(1-\pi_i)f_{S\mid Y,\boldsymbol{X}}(s_i\mid 0,\boldsymbol{x}_i)+\pi_if_{S\mid Y,\boldsymbol{X}}(s_i\mid 1,\boldsymbol{x}_i)p_{1i}\right\}^{\mathbbm{1}(s_i\leq 1)z_iw_i}\times\\
&\left\{(1-\pi_i)f_{S\mid Y,\boldsymbol{X}}(s_i\mid 0,\boldsymbol{x}_i)(1-p_{0i})+\pi_if_{S\mid Y,\boldsymbol{X}}(s_i\mid 1,\boldsymbol{x}_i)\right\}^{\mathbbm{1}(s_i>1)(1-z_i)w_i},
\end{aligned}\nonumber
\end{equation}
where 
$\boldsymbol{\eta}=(\eta_{0},\eta_{1})^\top$ and $\boldsymbol{\theta}=(\boldsymbol{\eta}^\top,\alpha,\xi_Y,\boldsymbol\xi^\top,\boldsymbol{\beta}^\top)^\top$.

Note that when $Y$ is observed for all subjects ($\mathbbm{1}(S_i\leq 1)\neq Z_i $ holds for $i=1,\cdots,n$), $\mathcal{L}(\boldsymbol{\theta})$ reduces to $\prod_{i=1}^n \left[\left\{\pi_if_{S\mid Y,\boldsymbol{X}}(s_i\mid 1,\boldsymbol{x}_i)(1-p_{1i})\right\}^{1-z_i}\left\{(1-\pi_i)f_{S\mid Y,\boldsymbol{X}}(s_i\mid 0,\boldsymbol{x}_i)p_{0i}\right\}^{z_i}\right]^{w_i}$, which is $\prod_{i=1}^n\{\pi_i^{1-z_i}(1-\pi_i)^{z_i}\}^{w_i}$ times a term independent of $\boldsymbol{\beta}$. The maximum likelihood estimates of $\boldsymbol{\beta}$ equal the estimates given by a naive logistic regression model. We employ the term ``naive logistic regression" to differentiate the basic logistic regression model of $Y$ on $\boldsymbol{X}$ using only data with observed $Y$, from the logistic regression component of our method (Equation~\ref{eq:pi_i}).

To account for the complex survey design, PHIA recommends using Jackknife (JK) repeated replication for variance estimation. PHIA surveys use a special case in which each stratum contains exactly two primary sampling units, a variation of the JK method known as JK2. The JK2 variance estimator of $\theta$ is $v_{\text{JK}2}=\sum_{k=1}^{H}\left(\widehat{\theta}_{(k)}-\widehat{\theta}\right)^{2}$, where $H$ is the number of strata, $\widehat{\theta}_{(k)}$ is the estimate of $\theta$ based on the $k$-th set of JK replicate weights, and $\widehat\theta$ is the estimate based on the full sample \citep{lu2006algorithms}.

Here, the main assumption that we have made in developing the model is that given $S$ and $Y$, the distribution of $Z$ does not depend on $\boldsymbol{X}$. This implies that the result of the last HIV test does not depend on age, gender, and biomarkers, given the test time and infection recency status. To check this assumption, we build a logistic regression model of $Z$ on $S$, $Y$, and $\boldsymbol{X}$. When $Y$ is missing, the predicted $Y$ from our model is used as a surrogate. A likelihood ratio test to compare this model to the reduced model which does not include $\boldsymbol{X}$ gives a p-value of 0.20.

\subsection{Semiparametric Likelihood Inference}
\label{subsec:extension}
As we will show in Section~\ref{sec:realdata}, we can assume $S\perp \boldsymbol{X}\mid Y$, i.e. $f_{S\mid Y,\boldsymbol{X}}(s\mid y,\boldsymbol{x})=f_{S\mid Y}(s\mid y)$. Instead of assuming the distribution type of $S\mid Y$, we assume a two-sample semiparametric density ratio model which links $f_{S\mid Y}(s\mid Y=1)$ and $f_{S\mid Y}(s\mid Y=0)$ by an exponential tilt $\exp(\psi_0+\psi_1s)$, following \cite{qin1994empirical}: $f_{S\mid Y}(s\mid Y=1)/f_{S\mid Y}(s\mid Y=0)=\exp(\psi_0+\psi_1s)$. This holds for two Gamma distributions with the same shape parameter.

We consider the non-parametric modeling for $f_{S\mid Y}(s\mid Y=0)$ following \cite{breslow1972comment}, in which the CDF of $S$ given $Y=0$ has a jump $p(s)$ at each $s$. Note that $s_i=s_j$ leads to $p(s_i)=p(s_j)$. Let $\boldsymbol{\theta}=(\boldsymbol{\eta}^\top,\psi_0,\psi_1,\boldsymbol{\beta}^\top)^\top$. The log likelihood function is then:
\begin{equation}
\begin{aligned}
l(\boldsymbol{\theta},p(s_1),\cdots,p(s_n))&=\sum_{i=1}^n w_i\left[\log\{p(s_i)\}+\mathbbm{1}(s_i\leq 1)(1-z_i)\log\{\pi_i\exp(\psi_0+\psi_1s_i)(1-p_{1i})\}+\right.\\
    &\mathbbm{1}(s_i>1)z_i\log\{(1-\pi_i)p_{0i}\}+\mathbbm{1}(s_i\leq 1)z_i\log\{1-\pi_i+\pi_i\exp(\psi_0+\psi_1s_i)p_{1i}\}+\\
    &\left.\mathbbm{1}(s_i>1)(1-z_i)\log\{(1-\pi_i)(1-p_{0i})+\pi_i\exp(\psi_0+\psi_1s_i)\}\right],
\end{aligned}\nonumber
\end{equation} under the constraints that $0\leq p(s)\leq 1$ for any $s\in\mathcal{S}$, $\sum_{s\in\mathcal{S}} p(s)=1$, $\sum_{s\in\mathcal{S}} p(s)\{\exp(\psi_0+\psi_1s)-1\}=0$, where $\mathcal{S}=\{s_i:1\leq i\leq n\}$.

We employ the Lagrange multiplier method to maximize the log likelihood \citep{qin1994empirical}. Let $H=l(\boldsymbol{\theta},p(s_1),\cdots,p(s_n))-\lambda\left\{\sum_{s\in\mathcal{S}} p(s)-1\right\}-n\mu\sum_{s\in\mathcal{S}} p(s)\{\exp(\psi_0+\psi_1s)-1\}$, where $\lambda,\mu$ are Lagrange multipliers. Taking derivative with respect to $p_s$, we have
\begin{equation}
\begin{aligned}
\frac{\partial H}{\partial p(s)}&=\frac{1}{p(s)}\sum_{\substack{1\leq i\leq n\\s_i=s}}w_i-\lambda-n\mu\{\exp(\psi_0+\psi_1s)-1\}=0,
    \\
    \sum_{s\in\mathcal{S}} \frac{\partial H}{\partial p(s)}p(s)&=\sum_{i=1}^n w_i-\lambda=0\quad\Rightarrow\quad \lambda=\sum_{i=1}^n w_i=n,\\
    p(s)&=\frac{\sum_{\substack{1\leq i\leq n\\s_i=s}}w_i}{n+n\mu\{\exp(\psi_0+\psi_1s)-1\}}.
\end{aligned}
\end{equation}
Note that $w_i$'s have been rescaled so that they add up to $n$. With the constraint that $\sum_{s\in\mathcal{S}} p(s)=1$, $\mu$ can be determined in terms of $\psi_0$, $\psi_1$. Also, from $0\leq p(s)\leq 1$ we derive $\max_{s\in\{s\in\mathcal{S}|u_s<0\}} u_s \leq \mu \leq \min_{s\in\{s\in\mathcal{S}|u_s>0\}} u_s$, where $u_s=(\sum_{\substack{1\leq i\leq n\\s_i=s}}w_i/n-1)\{\exp(\psi_0+\psi_1s)-1\}^{-1}$, $s\in\mathcal{S}$. Ignoring a constant, the profile log likelihood function for $\boldsymbol{\theta}$ is now defined as 
\begin{equation}
    \begin{aligned}
    l_p(\boldsymbol{\theta})&=\sum_{i=1}^n w_i\left[-\log\{1+\mu(\psi_0,\psi_1)(\exp(\psi_0+\psi_1s_i)-1)\}+\right.\\
    &\mathbbm{1}(s_i\leq 1)(1-z_i)\log\{\pi_i\exp(\psi_0+\psi_1s_i)(1-p_{1i})\}+\\
    &\mathbbm{1}(s_i>1)z_i\log\{(1-\pi_i)p_{0i}\}+\mathbbm{1}(s_i\leq 1)z_i\log\{1-\pi_i+\pi_i\exp(\psi_0+\psi_1s_i)p_{1i}\}+\\
    &\left.\mathbbm{1}(s_i>1)(1-z_i)\log\{(1-\pi_i)(1-p_{0i})+\pi_i\exp(\psi_0+\psi_1s_i)\}\right].
\end{aligned}
\label{eq:profile}
\end{equation}
We maximize this function with respect to $\boldsymbol{\theta}$ and then calculate the JK2 variance estimator.

\subsection{Prediction}

The proposed model enables the calculation of two types of risk to measure the probability of recent infection for each subject:

\paragraph{Type-1 risk: $P(Y=1\mid \boldsymbol{X}=\boldsymbol{x})\quad$ } 
Type-1 risk applies to all subjects. We can estimate Type-1 risk by simply replacing $\boldsymbol{\beta}$ with $\widehat{\boldsymbol{\beta}}$ in Equation~(\ref{eq:pi_i}).

\paragraph{Type-2 risk: $P(Y=1\mid S=s,Z=z,\boldsymbol{X}=\boldsymbol{x})\quad$ }

We have shown $P(Y=1\mid S=s,Z=0,\boldsymbol{X}=\boldsymbol{x})\equiv 1$ when $s\leq 1$, $P(Y=1\mid S=s,Z=1,\boldsymbol{X}=\boldsymbol{x})\equiv 0$ when $s>1$. Thus, Type-2 risk is meaningful only for subjects with $S\leq 1,Z=1$ or $S>1,Z=0$:

For $s\leq 1$,
\begin{equation}
\begin{aligned}
&P(Y=1 \mid S=s, Z=1, \boldsymbol{X}=\boldsymbol{x})
=\frac{P(Y=1 \mid \boldsymbol{X}=\boldsymbol{x}) f(s, Z=1 \mid Y=1, \boldsymbol{x})}{\sum_{j=0}^1  P(Y=j \mid \boldsymbol{X}=\boldsymbol{x}) f(s, Z=1 \mid Y=j, \boldsymbol{x})} \\
=&\left\{\frac{P(Y=0\mid \boldsymbol{X}=\boldsymbol{x})}{P(Y=1 \mid \boldsymbol{X}=\boldsymbol{x})}\cdot\frac{f(s\mid Y=0,\boldsymbol{x})}{f(s\mid Y=1,\boldsymbol{x})}\cdot\frac{1}{P(Z=1 \mid S=s, Y=1,\boldsymbol{X}=\boldsymbol{x})}+1 \right\}^{-1} \\
=&\left[\exp\left\{-\boldsymbol{x}^{\top} \boldsymbol\beta+\log f(s\mid Y=0,\boldsymbol{x})-\log f(s\mid Y=1,\boldsymbol{x})-\log p_1\right\}+1\right]^{-1}.
\end{aligned}
\end{equation}

For $s>1$,
\begin{equation}
\begin{aligned}
&P(Y=1 \mid S=s, Z=0, \boldsymbol{X}=\boldsymbol{x})
=\frac{P(Y=1 \mid \boldsymbol{X}=\boldsymbol{x}) f(s,Z=0 \mid Y=1, \boldsymbol{x})}{\sum_{j=0}^1 P(Y=j \mid \boldsymbol{X}=\boldsymbol{x}) f(s,Z=0 \mid Y=j, \boldsymbol{x})} \\
=&\left\{\frac{P(Y=0 \mid \boldsymbol{X}=\boldsymbol{x})}{P(Y=1 \mid \boldsymbol{X}=\boldsymbol{x})} \cdot\frac{f(s\mid Y=0,\boldsymbol{x})}{f(s\mid Y=1,\boldsymbol{x})}\cdot P(Z=0 \mid S=s, Y=0,\boldsymbol{X}=\boldsymbol{x})+1\right\}^{-1} \\
=&\left[\operatorname{exp}\left\{-\boldsymbol{x}^{\top} \boldsymbol{\beta}+\log f(s\mid Y=0,\boldsymbol{x})-\log f(s\mid Y=1,\boldsymbol{x})+\log (1-p_{0})\right\} +1\right]^{-1}.
\end{aligned}
\end{equation}

The HIV recency rate, $E(Y)$, is estimated at the weighted average of Type-2 risk for all individuals in our sample. Then for a given population, if external data sources are available to estimate ART coverage rate among HIV-positive adults ($P_\text{ART}$) and HIV prevalence ($P_\text{HIV}$), the estimation of annual HIV incidence can be obtained by
\begin{equation}
    \text{HIV incidence}=\frac{P_\text{HIV}\times(1-P_\text{ART})\times E(Y)}{(1-P_\text{HIV})+P_\text{HIV}\times(1-P_\text{ART})\times E(Y)},
   \label{eq:incidence}
\end{equation}
where the numerator is the proportion of recent infections in the population, and the denominator is the proportion of those at risk for new HIV infections in the past year. 

Equation~(\ref{eq:incidence}) assumes that individuals on ART are all long-term infections. While this assumption seems reasonable in the current data example, it may become problematic in more recent years as many countries have adopted the Universal Test and Treat (UTT) guideline. Under UTT, all individuals testing HIV positive are referred for immediate ART initiation, meaning the start of ART may occur closer to the time of infection. Our classification method remains valid for individuals not on ART, regardless of UTT implementation. However, to reliably estimate HIV incidence in the context of UTT, a separate recency classifier for individuals on ART will need to be developed, which we leave for future work.

\baselineskip=26pt

\section{HIV Recency Classification from PHIA Data}
\label{sec:realdata}

This section presents our model results on Malawi PHIA data. In Section~\ref{subsec:select}, we show our model selection procedures and final model results. We then compare the results to those obtained from the \textcolor{black}{naive} logistic regression that only uses observed values of $Y$ and the binary classification tree from the PHIA guideline in Section~\ref{subsec:modelcomparison}. 

\subsection{Comparison of Different Model Choices}
\label{subsec:select}

When fitting the full parametric model on Malawi PHIA data, it turns out that while $\xi_0$ has a p-value $<0.001$, all other effects corresponding to $\boldsymbol\xi$ in Equation~(\ref{eq:gamma}) have p-values $>0.30$. This inspires a reduced model with $\boldsymbol\xi=(\xi_0,0,0,0,0,0)^\top$. This reduced model achieves lower AIC and BIC, and a likelihood ratio test to compare it to the full model gives a p-value of 0.75 (see Web Table 2). Thus, Malawi PHIA data suggests a simplified model where we can assume $S\perp\boldsymbol X\mid Y$.

\subsection{Comparison with Naive Logistic Regression and RITA}
\label{subsec:modelcomparison}
In this subsection, we compare our model results to those obtained from the naive logistic regression as well as from RITA currently adopted by the PHIA surveys.

The naive logistic regression focuses on the subset of individuals whose recency status can be fully determined by the HIV testing history (the diagonal cells in Figure~\ref{fig1}). Table~\ref{tab:final} shows the coefficient estimates for our model (parametric and semiparametric) and the naive logistic regression model. The three models give consistent results: 
Age and ODn are negatively associated with recent infections, with the p-values being 0.02 and $<0.001$ respectively. The p-values for gender, CD4, and logVL are much larger than 0.05, suggesting their association with recent infection is weak. Among non-ART HIV-positive adults, younger people with lower ODn are more likely to be recent cases. There is no strong evidence that CD4 and logVL are associated with recent infection, though they are important indicators of HIV/AIDS progression \citep{shoko2019superiority}. One possible reason is that during the earliest stage of HIV infection, HIV viruses replicate rapidly, killing CD4 cells and promoting their migration from the blood into lymph nodes. As the immune system gains control over viral replication and establishes a steady state, viral load declines markedly, and CD4 cell counts partially recover before the eventual immune system failure without treatment \citep{le2021hitchhiker}.

Our method achieves smaller standard error than the naive logistic regression since we utilize more data. 
As for the HIV recency rate, our model gives lower estimate than naive logistic regression. It is because the naive logistic regression ignores the individuals with unknown recency status and assumes this response is missing at random. 
Web Figure 2 presents the distributions of biomarkers that are strong predictors of HIV recency status, and the distribution differs by whether the HIV recency status is known.

\begin{table}[!h]
\caption{Parameter estimates and standard error from our final model and naive logistic regression fitted on Malawi PHIA data.}
\begin{tabular*}{\hsize}{@{}@{\extracolsep{\fill}}lccc@{}}
\toprule
& Our model (parametric) & Our model (semiparametric) & Naive logistic regression\\
\midrule
$\alpha$ & $\phantom{-}1.05\;(0.10)$ &  & \\
$\xi_0$ & $-1.60\;(0.10)$ & &\\
$\xi_Y$ & $\phantom{-}1.80\;(0.17)$ & &\\
$\psi_0$ & & $\phantom{-}1.14\;(0.63)$ &\\
$\psi_1$ & & $-0.46\;(0.36)$ &\\
$\eta_{0}$&	$-0.74\;(0.09)$ & $-0.63\;(0.11)$&\\
$\eta_{1}$	&$\phantom{-}0.14\;(0.06)$	&  $\phantom{-}0.09\;(0.08)$&\\
$\beta_0$	&	$\phantom{-}0.12\;(0.23)$ & $\phantom{-}0.32\;(0.25)$& $\phantom{-}0.69\;(0.23)$ \\
$\beta_{\text{age}}$	& 	$-0.31\;(0.14)$ &	$-0.36\;(0.15)$& $-0.41\;(0.16)$\\
$\beta_{\text{gender}}$&		$-0.22\;(0.31)$ &	$-0.31\;(0.32)$& $-0.14\;(0.41)$\\
$\beta_{\text{CD4}}$	&	$-0.03\;(0.17)$ &	$\phantom{-}0.03\;(0.22)$& $-0.24\;(0.22)$\\
$\beta_{\text{ODn}}$	&	$-0.48\;(0.14)$ &	$-0.55\;(0.18)$ & $-0.68\;(0.20)$\\
$\beta_{\text{logVL}}$	&	$-0.12\;(0.15)$ &	$-0.15\;(0.16)$& $-0.19\;(0.21)$\\
$E(Y)$ & $\phantom{-}0.50\;(0.04)$ & $\phantom{-}0.53\;(0.05)$ & $\phantom{-}0.63\;(0.04)$\\
\bottomrule
\end{tabular*}
\label{tab:final}
\end{table}

We then examine the individual predicted probability of HIV recency status. For all individuals with or without known HIV recency status, Figure~\ref{fig2} compares the predictions using different methods. 
The naive logistic regression estimates the average probability of recent infection to be 0.67 for individuals with negative HIV testing results within one year ($Z=0$ and $S\leq1$) and 0.51 for individuals with positive HIV testing results before one year ($Z=1$ and $S>1$). In contrast, our model successfully classifies the former cases as recent infections and the latter as long-term infections with no ambiguity. Our Type-2 risk of recent infection diminishes to $0$ as $S$ approaches one when $Z=1$. When $Z=0$, the Type-2 risk is almost $0$ when $S\geq 5$, meaning that people whose latest HIV test results are negative are predicted to be most likely to have long-term infection if the test was over five years ago. 

\begin{figure}[!h]
    \centerline{\includegraphics[width=0.8\textwidth]{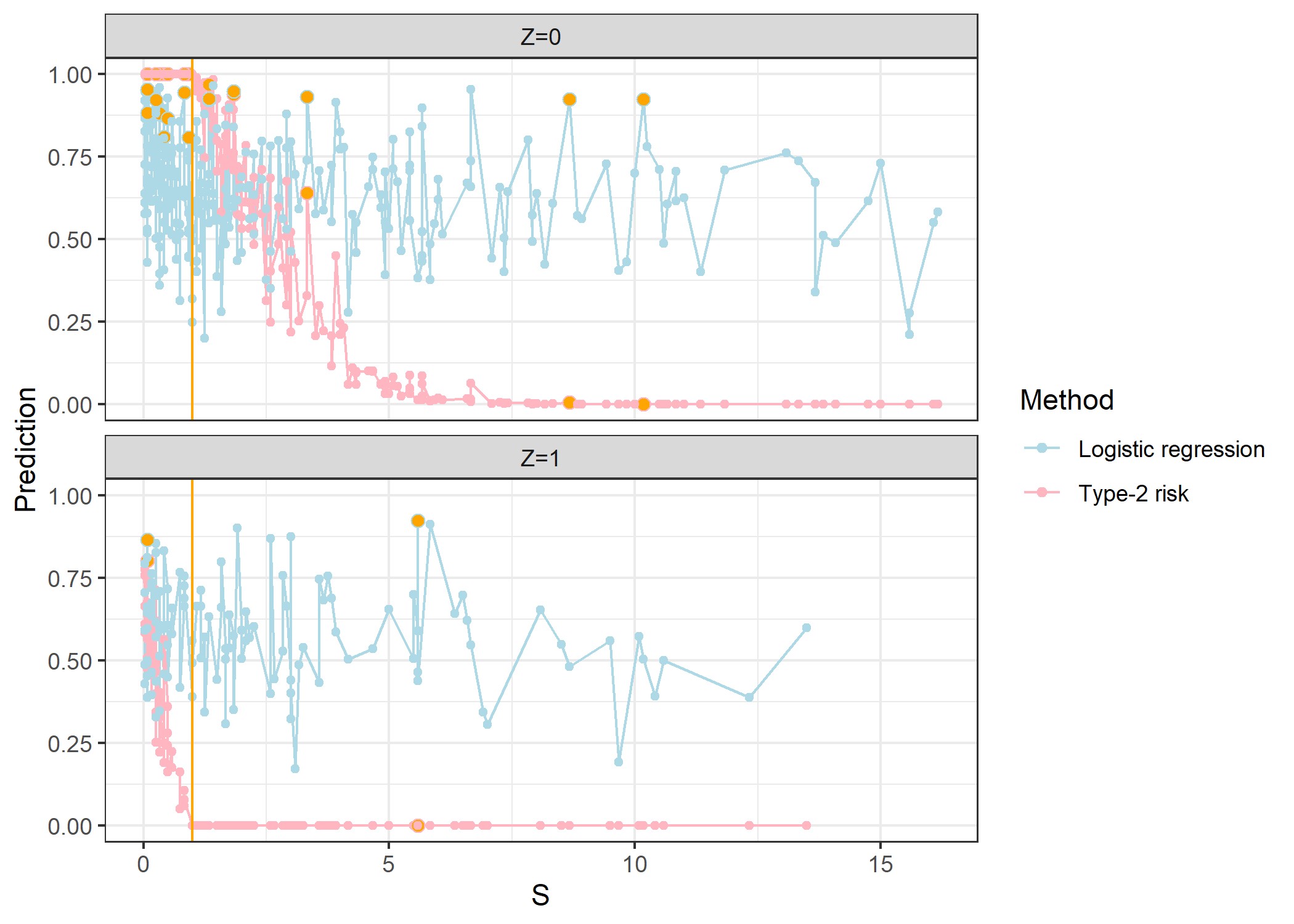}}
    \caption{Line charts of predicted response versus $S$ for Malawi PHIA individuals using naive logistic regression and Type-2 risk. Each dot represents a prediction of $Y$ for one individual using one method. Individuals classified as recent infection by the RITA are highlighted in orange. The orange vertical line marks $S=1$.}
    \label{fig2}
\end{figure}

In contrast with our method and naive logistic regression, RITA in PHIA does not give probabilistic predictions. It classifies a non-ART HIV-positive individual as a recent infection if ODn $\leq 1.5$ and VL $\geq1000$, and as long-term otherwise \citep{icap2021phia}. 
We set $\beta_{\text{age}}=\beta_{\text{gender}}=\beta_{\text{CD4}}=0$ in our model and only estimate $(\beta_{0},\beta_{\text{ODn}},\beta_{\text{logVL}})$ for biomarkers used by RITA. The estimates based on Malawi PHIA data are $(0.01,-0.49,-0.11)$ with standard error $(0.18,0.14,0.13)$. Lower ODn is related to recent infection, while logVL is likely not to be associated with infection recency.
Figure~\ref{fig2} shows that individuals classified as recent infection by RITA mostly have a large Type-2 risk except for three subjects whose most recent HIV test was taken over five years ago. 
Figure~\ref{fig3} compares the predictions from RITA and the Type-1 risk of recent infection. Among 408 Malawi PHIA individuals, only 17 are predicted as recent cases by RITA, all associated with higher Type-1 risk. Many solid circles are located outside the top-left quadrant, indicating low sensitivity of the RITA. Meanwhile, by our model, most individuals with long-term infection have a Type-1 risk lower than 0.45.

\begin{figure}[!h]
    \centerline{\includegraphics[width=\textwidth]{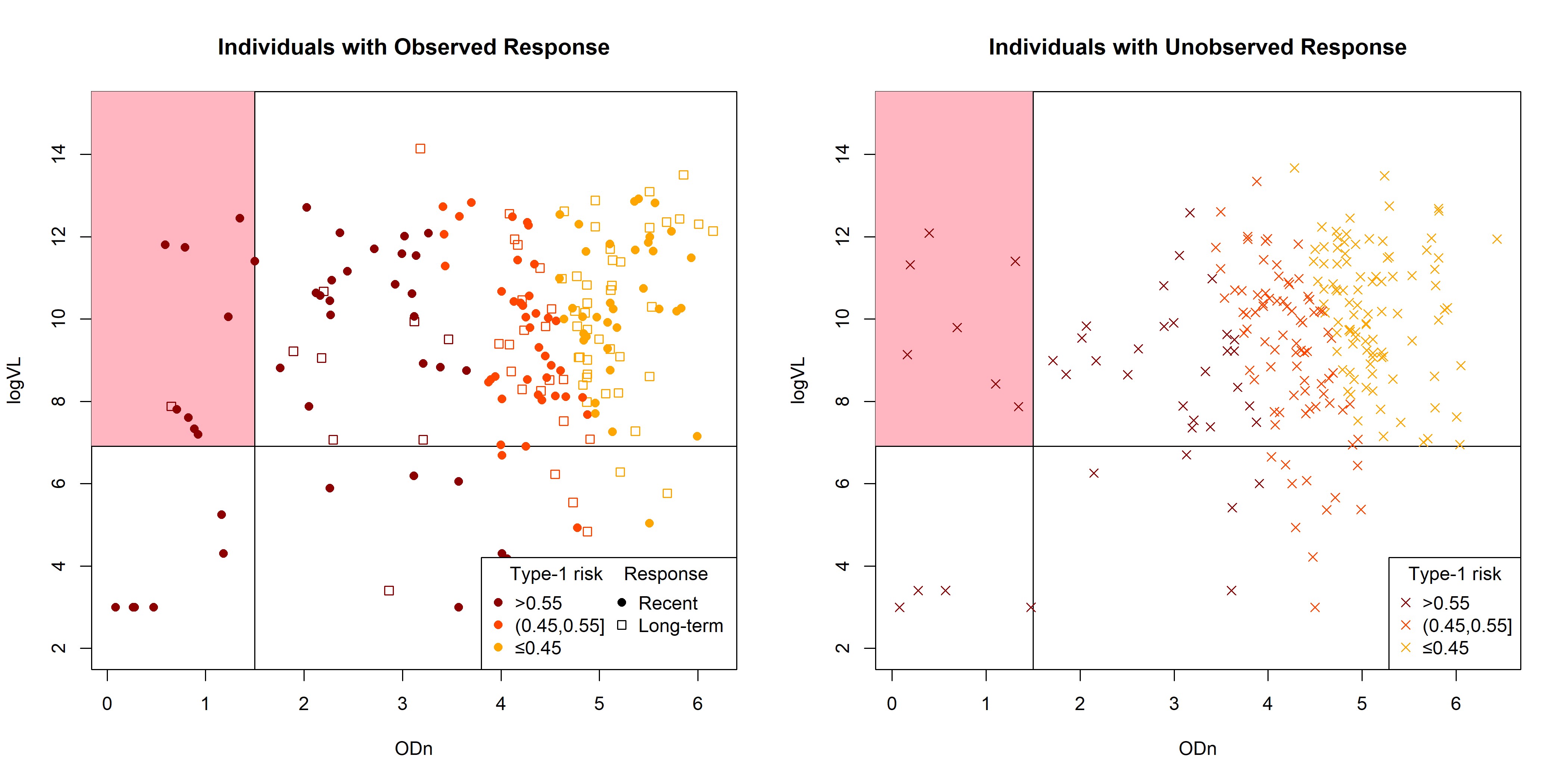}}
    \caption{Scatterplot of ODn and logVL for Malawi PHIA individuals. The top-left quadrant in each subplot represents recent infection predicted by RITA, and the other three quadrants represent long-term infection predicted by RITA. The color of the dots represents the Type-1 risk of recent infection predicted by our model. Based on $S$ and $Z$ values, some individuals have known HIV recency status, with solid circles representing recent infection and squares representing long-term infection. Crosses represent unknown HIV recency status.}
    \label{fig3}
\end{figure}

\section{Simulation Study}
\label{sec:sim}

Partially observed response renders it difficult to check the predictive performance of our model on real datasets. Therefore, we use a series of simulation studies to compare our model to naive logistic regression and the semi-supervised logistic regression model proposed by \cite{Sheng2023}, as well as to examine our model performance under potential reporting error and model misspecification. For simplicity, we consider subjects with equal sampling weights, and focus on the parametric version of our model from Simulation \Romannum{2} through \Romannum{7}.

\subsection{Comparison with Naive Logistic Regression}

Naive logistic regression only uses data with observed response $Y$, i.e. subjects with $S>1, Z=1$ or $S\leq1, Z=0$, while our method also incorporates data with unknown responses, as long as $S$ and $Z$ are observed. When data with observed responses barely represent the total population, naive logistic regression may lead to biased parameter estimates.

{\bf Simulation \Romannum{1}:} We demonstrate the above comparison through a simulation experiment, which takes the following steps:
\begin{enumerate}
\itemsep-0.35em 
\item Randomly generate the covariates ODn and age from standard normal distribution, and combine it with a column of $1$s to form the design matrix $\boldsymbol{X}$.

\item Simulate $Y$ using Equation~(\ref{eq:pi_i}), setting $\boldsymbol{\beta}$ to the estimate obtained from Malawi PHIA data.

\item Simulate $S$ using Equation~(\ref{eq:gamma}), setting $\alpha,\xi_0,\xi_Y$ to their estimates obtained from Malawi PHIA data.

\item Simulate $Z$ given $Y$ and $S$ using Equation~(\ref{eq:p0i_p1i}), with $\eta_{0},\eta_{1}$ set to their estimates obtained from Malawi PHIA data.

\item Randomly divide all observations into two groups (data.phia and data.test) of equal sample size. Then divide data.phia into two groups: data.phiaA with observed responses; data.phiaB with unobserved responses. Similarly, divide data.test into two groups: data.testA with observed responses; data.testB with unobserved responses.

\item Fit a naive logistic regression model on data.phiaA, and make predictions on data.test. Perform our method (parametric and semiparametric) on data.phia, and then calculate Type-1 risk, $P(Y_i=1\mid \boldsymbol{X}_i=\boldsymbol{x}_i)$, for data.test and Type-2 risk, $P(Y_i=1\mid S_i=s_i,Z_i=z_i,\boldsymbol{X}_i=\boldsymbol{x}_i)$, for data.testB. 
\end{enumerate}

Table~\ref{tab:sim1} compares the results between the three models. They give similar estimates of $\beta_{\text{ODn}}$ and $\beta_{\text{age}}$. As expected, our method achieves smaller standard errors since it uses more training data by including the observations with unknown responses. Naive logistic regression gives biased estimates of the intercept parameter $\beta_0$ and $E(Y)$. While predictions from the naive logistic regression and our Type-1 risk from both parametric and semiparametric model all achieve an average AUC of $0.65$ on data.test, predictions from Type-2 risk achieve an average AUC of 0.92 on data.testB, showing the benefit of using HIV testing history in our model.

\begin{table}[!h]
\caption{Average parameter estimate, average standard error, standard deviation, and coverage probability of 95\% confidence interval over 500 replicates. The training sample size of our method is 1000 while, on average, 451 of them have an observed response. Test sample size is 1000.}
\small
\centering
\begin{tabular*}{\hsize}{@{}@{\extracolsep{\fill}}lcccccccccc@{}}
\toprule
& & \multicolumn{3}{c}{Our method (parametric)} & \multicolumn{3}{c}{Our method (semiparametric)} & \multicolumn{3}{c}{Naive logistic regression}\\
\cmidrule{3-5}
\cmidrule{6-8}
\cmidrule{9-11}
 & true value & estimate (se) & sd & coverage & estimate (se) & sd & coverage & estimate (se) & sd & coverage \\ 
\midrule
$\alpha$ & $\phantom{-}1.07$& $\phantom{-}1.08\;(0.06)$ & 0.06 & 0.95 & & &&&&\\
$\xi_0$ & $-1.59$ & $-1.58\;(0.07)$ & 0.07 & 0.94 & & &&&&\\
$\xi_Y$ & $\phantom{-}1.83$ & $\phantom{-}1.82\;(0.08)$ & 0.08 & 0.96  & & &&&&\\
$\psi_{0}$ & $\phantom{-}1.96$ &	 &	&	 & $\phantom{-}2.00\;(0.28)$  & 0.28 &0.96 &   &  &  \\ 
$\psi_{1}$ & $-1.07$&	&	&	& $-1.11\;(0.22)$  & 0.22 & 0.95 &   &  &  \\ 
$\eta_{0}$ & $-0.74$ &	$-0.74\;(0.06)$ &	$0.06$&	$0.94$ & $-0.75\;(0.06)$  & 0.06 &0.95 &   &  &  \\ 
$\eta_{1}$ & $\phantom{-}0.15$&	$\phantom{-}0.15\;(0.03)$ &	$0.03$&	$0.95$ & $\phantom{-}0.15\;(0.03)$  & 0.03 & 0.96 &   &  &  \\ 
$\beta_0$ & $\phantom{-}0.02$&	$\phantom{-}0.03 \;(0.09)$ &	$0.09$ &	0.96&	$\phantom{-}0.02\;(0.09)$ & 0.08  & 0.96 & $\phantom{-}0.63 \;(0.10)$&	0.10&	0\\
$\beta_{\text{ODn}}$ & $-0.50$&	$-0.51 \;(0.08)$&	$0.08$ &	0.96&	$-0.51\;(0.08)$ & 0.08  & 0.97 & $-0.51 \;(0.11)$&	0.11&	0.95\\ 
$\beta_{\text{age}}$ & $-0.29$&	$-0.29 \;(0.08)$&	$0.08$ &	0.95&	$-0.29\;(0.08)$ & 0.08  & 0.94 & $-0.29 \;(0.11)$&	0.11&	0.94\\ 
$E(Y)$ & $\phantom{-}0.50$ & $\phantom{-}0.51$ & $0.02$ &  & 0.50 & 0.02  &  & $\phantom{-}0.64$ & 0.02 & \\
\bottomrule
\end{tabular*}
\label{tab:sim1}
\end{table}

\subsection{Comparison with Semi-Supervised Logistic Regression}

\cite{Sheng2023} proposed a semi-supervised logistic regression model that combines the PHIA dataset with external contingency tables from cohort studies, which show the number of specimens in covariate groups given true HIV recency status. The relationship between $\boldsymbol{X}$ and $Y$ is assumed to be the same across those datasets. They do not consider self-reported HIV testing history and treat HIV recency status as completely unknown in PHIA data. Simulations \Romannum{2}–\Romannum{4} in Web Appendix B compare our method and Sheng's method.

\subsection{Sensitivity  Analysis}
Simulations \Romannum{5}–\Romannum{7} in Web Appendix C demonstrate our model's robustness against reporting errors and model misspecifications. Specifically, Simulation \Romannum{5} considers reporting error in $Z$; Simulation \Romannum{6} considers telescoping bias in $S$; Simulation \Romannum{7} investigates what will happen when $p_0, p_1$ depend on covariates in $\boldsymbol{X}$ but we mistakenly assume $Z\perp\boldsymbol{X}\mid S,Y$.

\section{Discussion}
\label{sec:discuss}

In this paper, we proposed a likelihood-based probabilistic model for HIV recency classification using self-report HIV testing history and biomarker data collected by the PHIA Project. The utility of self-report testing history in HIV incidence estimation has garnered attention in recent literature. For example, \cite{Fellows2023jv} used the self-reported time between the last HIV test and the survey to estimate the time between infection and diagnosis. We demonstrate that self-report testing history is also useful in HIV recency classification: HIV recency status could be determined for specific HIV testing history, turning HIV recency classification into a semi-supervised learning problem with partially observed responses. The observed data likelihood integrates the mechanism of how HIV recency status depends on biomarkers and the mechanism of how HIV recency status, together with the self-report time of the last HIV test, impacts the self-report test results.

We predicted HIV recency status for Malawi PHIA data and compared our method to the naive logistic regression model that only uses data with observed response and the binary classification tree (current practice).
Our method achieved smaller standard error than naive logistic regression, and captured more recent cases than the binary classification tree. To check the performance for individuals with unknown HIV recency status, we designed a series of simulation studies where datasets were generated to mimic the observed Malawi PHIA data. Our Type-2 risk of recent infection, which uses the self-report HIV testing history beyond biomarkers, is shown to be a better prediction of HIV recency status than Type-1 risk relying merely on biomarkers. Our estimate of the HIV recency rate based on Type-2 risk is relatively robust to potential reporting error and model misspecification. Moreover, since our method incorporates both individuals with known recency status based on their testing histories
and individuals whose recency status could not be determined, we can obtain more efficient and less biased parameter estimates than naive logistic regression. Naive logistic regression is a complete-case analysis, where the subjects with observed $Y$ come from a biased sample. Particularly, the missingness of $Y$ is determined by the combination of $S$ and $Z$ by definition, while the correlation of $Y$ and $Z$ is governed by the assumptions on $p_{0i}$ and $p_{1i}$. Consequently, $Y$ is correlated with its missingness indicator, making the complete cases a biased sample.

We proposed a parametric model where we assumed Gamma distribution for $S\mid Y,\boldsymbol{X}$. To allow more flexibility, we also proposed a semiparametric exponential tilting model as an alternative to estimate the distribution of $S\mid Y$, after making the assumption of $S\perp\boldsymbol X\mid Y$ on Malawi PHIA data. Another implicit assumption is that the self-reported HIV testing history has negligible reporting error. A recent study shows that self-reported HIV status has very high ($>99\%$) specificity and high ($>85\%$) sensitivity \citep{chasimpha2020assessing}. The reporting error in time since most recent HIV test is ignored in studies such as \cite{fellows2020}. Simulation study also demonstrates robustness of our proposed model against a moderate level of reporting error. While it is interesting to extend our model to allow for self-reporting error, we would leave this for future exploration. 

\backmatter

\section*{Acknowledgements}
 Wenlong Yang, Le Bao and Runze Li were supported by NIH/NIAID under grants R01AI136664 and R01AI170249. Danping Liu's research was supported by the Intramural Research Program of the National Cancer Institute, the National Institutes of Health (NIH). We appreciate the PHIA teams making the data publicly available for research purposes. Authors are also thankful to the editor, associate editor and two anonymous reviewers for their constructive suggestions and comments.

\FloatBarrier

\section*{Supplementary Materials}
Web Appendices, Tables, Figures, and code referenced in Sections~\ref{sec:Data}, \ref{sec:Method}, \ref{sec:sim} are available with this paper at the Biometrics website on Oxford Academic. R code (.Rmd file) to perform the method described in this article covers real data analysis and simulation studies.

\section*{Data Availability}
The data that support the findings in this paper can be shared on reasonable request to the Population-based HIV Impact Assessment (PHIA) team. Data request instructions are provided in Supplementary Materials. A tutorial on accessing PHIA documentation and datasets can be found at \url{https://phia.icap.columbia.edu/analyzing-phia-data-in-stata/}.

\bibliographystyle{biom}
\bibliography{biblio}

\begin{thebibliography}{}

\bibitem[\protect\citeauthoryear{Breslow}{Breslow}{1972}]{breslow1972comment}
Breslow, N. (1972).
\newblock Comment on ``{R}egression and life tables" by {DR Cox}.
\newblock {\em Journal of the Royal Statistical Society, Series B} {\bf 34,} 216--217.

\bibitem[\protect\citeauthoryear{Chasimpha, Mclean, Dube, McCormack, dos Santos-Silva, and Glynn}{Chasimpha et~al.}{2020}]{chasimpha2020assessing}
Chasimpha, S.~J., Mclean, E.~M., Dube, A., McCormack, V., dos Santos-Silva, I., and Glynn, J.~R. (2020).
\newblock Assessing the validity of and factors that influence accurate self-reporting of {HIV} status after testing: a population-based study.
\newblock {\em AIDS (London, England)} {\bf 34,} 931.

\bibitem[\protect\citeauthoryear{Fellows, Hladik, Eaton, Voetsch, Parekh, and Shiraishi}{Fellows et~al.}{2023}]{Fellows2023jv}
Fellows, I.~E., Hladik, W., Eaton, J.~W., Voetsch, A.~C., Parekh, B.~S., and Shiraishi, R.~W. (2023).
\newblock Improving biomarker-based {HIV} incidence estimation in the treatment era.
\newblock {\em Epidemiology} {\bf 34,} 353--364.

\bibitem[\protect\citeauthoryear{Fellows, Shiraishi, Cherutich, Achia, Young, and Kim}{Fellows et~al.}{2020}]{fellows2020}
Fellows, I.~E., Shiraishi, R.~W., Cherutich, P., Achia, T., Young, P.~W., and Kim, A.~A. (2020).
\newblock A new method for estimating {HIV} incidence from a single cross-sectional survey.
\newblock {\em PLoS One} {\bf 15,} 1--12.

\bibitem[\protect\citeauthoryear{{ICAP at Columbia University} et~al\mbox{.}}{{ICAP at Columbia University} et~al.}{2021}]{icap2021phia}
{ICAP at Columbia University} et~al. (2021).
\newblock {Population-based HIV Impact Assessment (PHIA)} data use manual.
\newblock \url{https://phia-data.icap.columbia.edu/storage/Country/28-09-2021-22-01-17-615390ad6a147.pdf} (accessed April 18, 2024).

\bibitem[\protect\citeauthoryear{{Joint United Nations Programme on HIV/AIDS and the World Health Organization}}{{Joint United Nations Programme on HIV/AIDS and the World Health Organization}}{2022}]{world2022using}
{Joint United Nations Programme on HIV/AIDS and the World Health Organization} (2022).
\newblock Using recency assays for {HIV} surveillance: 2022 technical guidance.
\newblock \url{https://www.unaids.org/en/resources/documents/2023/using-recency-assays-HIV-surveillance} (accessed April 18, 2024).

\bibitem[\protect\citeauthoryear{Kassanjee, McWalter, B{\"a}rnighausen, and Welte}{Kassanjee et~al.}{2012}]{kassanjee2012new}
Kassanjee, R., McWalter, T.~A., B{\"a}rnighausen, T., and Welte, A. (2012).
\newblock A new general biomarker-based incidence estimator.
\newblock {\em Epidemiology} {\bf 23,} 721--728.

\bibitem[\protect\citeauthoryear{Laeyendecker, Brookmeyer, Cousins, Mullis, Konikoff, Donnell, et~al\mbox{.}}{Laeyendecker et~al.}{2013}]{laeyendecker2013hiv}
Laeyendecker, O., Brookmeyer, R., Cousins, M.~M., Mullis, C.~E., Konikoff, J., Donnell, D., et~al. (2013).
\newblock {HIV} incidence determination in the {United States}: a multiassay approach.
\newblock {\em The Journal of Infectious Diseases} {\bf 207,} 232--239.

\bibitem[\protect\citeauthoryear{Le~Hingrat, Sereti, Landay, Pandrea, and Apetrei}{Le~Hingrat et~al.}{2021}]{le2021hitchhiker}
Le~Hingrat, Q., Sereti, I., Landay, A.~L., Pandrea, I., and Apetrei, C. (2021).
\newblock The hitchhiker guide to {CD4+ T-cell} depletion in lentiviral infection. {A} critical review of the dynamics of the {CD4+ T cells in SIV and HIV infection}.
\newblock {\em Frontiers in Immunology} {\bf 12,} 695674.

\bibitem[\protect\citeauthoryear{Lu, Brick, and Sitter}{Lu et~al.}{2006}]{lu2006algorithms}
Lu, W.~W., Brick, J.~M., and Sitter, R.~R. (2006).
\newblock Algorithms for constructing combined strata variance estimators.
\newblock {\em Journal of the American Statistical Association} {\bf 101,} 1680--1692.

\bibitem[\protect\citeauthoryear{{Ministry of Health, Malawi, Centers for Disease Control and Prevention, and ICAP at Columbia University}}{{Ministry of Health, Malawi, Centers for Disease Control and Prevention, and ICAP at Columbia University}}{2016}]{mphia2016}
{Ministry of Health, Malawi, Centers for Disease Control and Prevention, and ICAP at Columbia University} (2016).
\newblock {MPHIA} 2015-2016 household interview and biomarker datasets.
\newblock \url{https://phia-data.icap.columbia.edu/datasets?country_id=3&year_id=2015} (accessed March 16, 2022).

\bibitem[\protect\citeauthoryear{Qin and Lawless}{Qin and Lawless}{1994}]{qin1994empirical}
Qin, J. and Lawless, J. (1994).
\newblock Empirical likelihood and general estimating equations.
\newblock {\em The Annals of Statistics} {\bf 22,} 300--325.

\bibitem[\protect\citeauthoryear{Sheng, Li, Bao, and Li}{Sheng et~al.}{2023}]{Sheng2023}
Sheng, B., Li, C., Bao, L., and Li, R. (2023).
\newblock {Probabilistic {HIV} recency classification—a logistic regression without labeled individual level training data}.
\newblock {\em The Annals of Applied Statistics} {\bf 17,} 108 -- 129.

\bibitem[\protect\citeauthoryear{Shoko and Chikobvu}{Shoko and Chikobvu}{2019}]{shoko2019superiority}
Shoko, C. and Chikobvu, D. (2019).
\newblock A superiority of viral load over {CD4} cell count when predicting mortality in {HIV} patients on therapy.
\newblock {\em BMC Infectious Diseases} {\bf 19,} 169.

\bibitem[\protect\citeauthoryear{Stirrup, Sabin, Phillips, Williams, Churchill, Tostevin, et~al\mbox{.}}{Stirrup et~al.}{2019}]{stirrup2019associations}
Stirrup, O.~T., Sabin, C.~A., Phillips, A.~N., Williams, I., Churchill, D., Tostevin, A., et~al. (2019).
\newblock Associations between baseline characteristics, {CD4} cell count response and virological failure on first-line efavirenz + tenofovir + emtricitabine for {HIV}.
\newblock {\em Journal of Virus Eradication} {\bf 5,} 204--211.

\bibitem[\protect\citeauthoryear{{UNAIDS}}{{UNAIDS}}{2023}]{aidsinfo2023}
{UNAIDS} (2023).
\newblock {AIDS}info.
\newblock \url{https://aidsinfo.unaids.org/} (accessed {M}arch 14, 2023).

\bibitem[\protect\citeauthoryear{Voetsch, Duong, Stupp, Saito, McCracken, Dobbs, et~al\mbox{.}}{Voetsch et~al.}{2021}]{voetsch2021hiv}
Voetsch, A.~C., Duong, Y.~T., Stupp, P., Saito, S., McCracken, S., Dobbs, T., et~al. (2021).
\newblock {HIV}-1 recent infection testing algorithm with antiretroviral drug detection to improve accuracy of incidence estimates.
\newblock {\em Journal of Acquired Immune Deficiency Syndromes} {\bf 87,} S73--S80.

\end{thebibliography}


\end{document}


\if0\blind
{
  \title[Supplementary Materials]{\bf Supplementary Materials for ``A Likelihood Approach to Incorporating Self-Report Data in HIV Recency Classification'' by Wenlong Yang, Danping Liu, Le Bao, Runze Li}
  \maketitle
} \fi

\if1\blind
{
  \bigskip
  \bigskip
  \bigskip
  \begin{center}
    {\LARGE\bf Supplementary Materials for ``A Likelihood Approach to Incorporating Self-Report Data in HIV Recency Classification''}
\end{center}
  \medskip
} \fi

\section{Alternative Link Functions for $p_{0i},p_{1i}$}
\label{sec:link}

Apart from the log link function in Equation~(\ref{eq:p0i_p1i}), we can also use other link functions to model $Z_i$, as long as they satisfy the continuity assumption, i.e. $\lim\limits_{S_i \downarrow 1}p_{0i}=1$ and $\lim\limits_{S_i \uparrow 1} p_{1i}=0$. The continuity assumption on $p_{0i}$ and $p_{1i}$ implies that when a subject with long-term infection received a test just over one year ago, the test result must have very high (close to 1) probability of being positive, and conversely, when a subject with recent infection received a test just within one year, the test result
must have very high probability of being negative.

For example, alternative options for link functions include logit link function
\begin{equation}
\begin{aligned}
    p_{0i}&=\operatorname{expit}(\eta_{00}+\eta_{01}\log s_i),\quad s_i>1,\\
    p_{1i}&=\operatorname{expit}(\eta_{10}+\eta_{11}\log s_i),\quad s_i\leq 1,
\end{aligned}
\end{equation}
where $\eta_{00}\gg 0$, $\eta_{10}\ll 0$, $\eta_{01},\eta_{11}<0$, and Weibull survival function
\begin{equation}
\begin{aligned}
    p_{0i}&=\exp\left\{-\left(\frac{\log s_i}{\lambda_0}\right)^{k_0}\right\},\quad s_i>1,\\
    p_{1i}&=1-\exp\left\{-\left(\frac{\log s_i}{\lambda_1}\right)^{k_1}\right\},\quad s_i\leq 1,
    \label{eq:weibull}
\end{aligned}
\end{equation}
where $\lambda_1<0$, $\lambda_0,k_0,k_1>0$. We assume the same shape parameter for $p_{0i}$ and $p_{1i}$, i.e. $k_0=k_1$ for simplicity. Equation~(\ref{eq:weibull}) can be viewed as a generalization of Equation~(\ref{eq:p0i_p1i}). {Web Table}~\ref{tab:etaselect} shows that on Malawi PHIA data, the smallest AIC and BIC are achieved by log link function, which is our preferred choice.

\section{Simulation Studies to Compare Our Method with Semi-Supervised Logistic Regression}
\label{sec:sheng}

Provided that Sheng's model is correctly specified, both our method and Sheng's method should give consistent coefficient estimates. We now compare them under three different simulation settings. 

{\bf Simulation \Romannum{2}:} We aim to compare the efficiency of Sheng's method and our method when Sheng's model uses a contingency table that is a good representation of the population and when the population HIV recency rate is around 50\%. We use unstandardized covariates logVL and ODn, in consistency with \cite{Sheng2023}. The simulation takes the following steps:
\begin{enumerate}
\itemsep-0.15em 
    \item Assume logVL and ODn follow a bivariate normal distribution and estimate the joint distribution using Malawi PHIA data. Randomly generate logVL and ODn from the fitted distribution, and combine them with a column of 1s to form the design matrix $\boldsymbol{X}$. Sample size $k=3000$.
    \item Simulate $Y$ based on $\pi_i:=P(Y_i=1\mid \boldsymbol{X}_i=\boldsymbol{x}_i)=\beta_0+\beta_1\text{logVL}_i+\beta_2\text{ODn}_i$, with $\beta_1$, $\beta_2$ set to their estimates obtained from Malawi PHIA data. Choose $\beta_0$ so that the average of $Y$ is around 50\%.
    \item Simulate $S$ using Equation~(\ref{eq:gamma}), setting $\alpha,\xi_0,\xi_Y$ to their estimates obtained from Malawi PHIA data. 
    \item Simulate $Z$ given $Y$ and $S$, using Equation~(\ref{eq:p0i_p1i}), with $\boldsymbol{\eta}$ set to the estimates obtained from Malawi PHIA data.
    \item Randomly divide all observations into three groups (data.contingency.table, data.phia, and data.test) with an equal sample size of 1000. Divide data.test into two groups: data.testA with observed responses; data.testB with unobserved responses.
    \item Based on data.contingency.table, construct a contingency table of viral load and ODn grouped by $Y$, with cut points determined according to quantiles.
    \item Perform our method on data.phia, and calculate Type-1 risk for data.test and Type-2 risk for data.testB. Perform Sheng's method on the contingency table and data.phia with $Y$ removed, and make predictions for data.test.
\end{enumerate}

The simulation results are presented in {Web Table}~\ref{tab:sim2}. The two methods obtain similar parameter estimates close to the true value, and the standard error is also similar since the training sample size of our method equals the sample size of the contingency table used in Sheng's model. While predictions from Sheng's model and Type-1 risk achieve an average AUC of $0.64$ on data.test, predictions from Type-2 risk achieve an average AUC of 0.92 on data.testB.

{\bf Simulation \Romannum{3}:} When Sheng's model uses a contingency table that misses important covariates in the primary dataset, we follow almost the same steps as in Simulation \Romannum{2}, but construct a univariate contingency table in Step 6. {Web Table}~\ref{tab:sim3} compares our method with Sheng's model using a univariate contingency table. In Sheng's model, when ODn or viral load is absent in the contingency table, the corresponding effect cannot be estimated, and estimates of other parameters are biased, except for the HIV recency rate. The missingness of ODn is worse than the missingness of viral load. The average AUC of Sheng's model on data.test is 0.56 when ODn is missing in the contingency table, and is 0.64 when viral load is missing in the contingency table.

{\bf Simulation \Romannum{4}:} When the contingency table used by Sheng's model does not represent the target population, we revise Step 6 of Simulation \Romannum{2} by regenerating $Y$ for data.contingency.table using $\pi_i:=P(Y_i=1\mid \boldsymbol{X}_i=\boldsymbol{x}_i)=\gamma_0+\gamma_1\text{logVL}_i+\gamma_2\text{ODn}_i$, with $\gamma_1$, $\gamma_2$ set to their estimates obtained from Zimbabwe PHIA data and $\gamma_0$ set to $\beta_0$ in Simulation \Romannum{2}. Based on the updated data.contingency.table, construct a contingency table of viral load and ODn grouped by $Y$, with cut points determined according to quantiles.
{Web Table}~\ref{tab:sim4} shows the simulation results. The estimates of $\boldsymbol{\beta}$ and $E(Y)$ obtained by Sheng's model are far from true values yet close to those used to generate the contingency table. The average AUC of Sheng's model on data.test drops to 0.60.

\section{Simulation Studies for Sensitivity Analysis}
\label{sec:sens}

Here, we use simulation analysis to evaluate the model performance under the presence of reporting errors and model misspecifications. Specifically, Simulation \Romannum{5} considers reporting error in $Z$; Simulation \Romannum{6} considers telescoping bias in $S$; Simulation \Romannum{7} investigates what will happen when $p_0, p_1$ depend on covariates in $\boldsymbol{X}$ but we mistakenly assume $Z\perp\boldsymbol{X}\mid S,Y$.

{\bf Simulation \Romannum{5}:} We examine how our method performs in comparison to the naive logistic regression when $Z$ is subject to reporting error. We follow the same steps as in Simulation \Romannum{1} except that right before Step 5, $Z$ is replaced with its noisy self-reported version $Z^*$ generated according to $P(Z^*=1\mid Z=1)=0.85$, $P(Z^*=0\mid Z=0)=0.99$. Thereby the self-reported $Z$ has a sensitivity of 0.85 and specificity of 0.99, matching the findings from a recent study \citep{chasimpha2020assessing}. As shown in Web Table~\ref{tab:sim5}, reporting error of $Z$ mainly affects the estimation of $\boldsymbol\eta$ and $\beta_0$. Estimation of $\beta_{\text{ODn}},\beta_{\text{age}}$ is less affected because the misclassification of $Z$ is independent of ODn and age. The prediction of HIV recency status is also less affected. The average AUC based on Type-1 risk remains 0.65 on data.test; the average AUC based on Type-2 risk attains 0.89 on data.testB; our method slightly overestimates $E(Y)$ but still outperforms the naive logistic regression.

{\bf Simulation \Romannum{6}:} We now examine the effects of reporting error in $S$. We consider forward telescoping bias (people tend to recall that distant events occurred more recently than is actually the case) and backward telescoping bias (people tend to recall that recent events occurred farther back in time) separately. We follow the same steps as in Simulation \Romannum{1} except that right before step 5, we apply a multiplicative noise uniformly distributed over $[0.7, 1.1]$ for $S>1$ (when we consider forward telescoping), or a multiplicative noise uniformly distributed over $[0.9, 1.3]$ for $S\leq 1$ (when we consider backward telescoping). Thereby the reported $S$ tends to deflate when the actual $S$ is large (or inflate when the actual $S$ is small). For example, when $S=5$ years, the error-prone reported $S$ is between 3.5 and 5.5 years.

As shown in Web Table~\ref{tab:sim6}, telescoping bias of $S$ mainly affects the estimation of the model of $S\mid Y$ and $\beta_0$, while the estimation of $\beta_{\text{ODn}},\beta_{\text{age}}$ is less affected, since the reporting error of $S$ is independent of age and ODn. The prediction of HIV recency status is also less affected. In either forward telescoping or backward telescoping, we observe that the average AUC based on Type-1 risk remains 0.65 on data.test; the average AUC based on Type-2 risk attains 0.92 on data.testB; $E(Y)$ is slightly overestimated.

{\bf Simulation \Romannum{7}:} 
We investigate what will happen when $p_0, p_1$ depend on covariates in $\boldsymbol{X}$ but we mistakenly assume the conditional independence of $Z$ and $\boldsymbol{X}$ given $S$ and $Y$. We follow almost the same steps as in Simulation \Romannum{1}, but in Step 4, we use $p_{0i}=\exp\{\exp(-\text{ODn}_i/2)\eta_0\log s_i\}$ and $p_{1i}=1-\exp\{\exp(\text{ODn}_i/2)\eta_1\log s_i\}$ instead. Web Table~\ref{tab:sim7} shows the simulation results. Most parameter estimates are biased, especially for $\beta_{\text{ODn}}$ and $\eta_0$. The prediction of HIV recency status is relatively robust to this type of model specification, though: The average AUC based on Type-1 risk remains 0.65 on data.test; the average AUC based on Type-2 risk attains 0.90 on data.testB; the estimated HIV recency rate is close to the true value. If we suspect that $p_0,p_1$ also depend on covariates in $\boldsymbol{X}$ when validating model assumptions, we should include the corresponding covariates in Equation~(\ref{eq:p0i_p1i}).

\newpage

\begin{table}[!ht]
\centering
\caption{Comparison of log likelihood (LL), AIC, and BIC across models using different link functions.}
\begin{tabular*}{\hsize}{@{}@{\extracolsep{\fill}}lccc@{}}
\toprule
& LL & AIC& BIC\\
\midrule
Log link function 	& $-1051.42$	& 2124.84& 	2168.96\\
Logit link function & $-1054.03$ & 2134.06 & 2186.20\\
Weibull survival function	& $-1051.40$	& $2126.81$& 	 $2174.94$\\
\bottomrule
\end{tabular*}
\label{tab:etaselect}
\end{table}

\begin{table}[!ht]
\centering
\caption{Comparison of log likelihood (LL), AIC, and BIC between the full model and the reduced model.}
\begin{tabular*}{\hsize}{@{}@{\extracolsep{\fill}}lccc@{}}
\toprule
& LL & AIC & BIC\\
\midrule
Full model & $-1050.08$ &  $2132.16$ &  $2196.34$ \\
Reduced model & $-1051.42$ &  2124.84  & 2168.96 \\
\bottomrule
\end{tabular*}
\label{tab:lrt}
\end{table}

\begin{table}[!ht]
\caption{Average parameter estimate, average standard error, standard deviation, and coverage probability of 95\% confidence interval over 500 replicates. Training sample size = test sample size = contingency table sample size = 1000.}
\centering
\begin{tabular*}{\hsize}{@{}@{\extracolsep{\fill}}lccccccc@{}}
\toprule
& & \multicolumn{3}{c}{Our method} & \multicolumn{3}{c}{Sheng's method}\\
\cmidrule{3-5}
\cmidrule{6-8}
& true value & estimate (se) & sd & coverage & estimate (se) & sd & coverage \\ 
\midrule
$\alpha$ & $\phantom{-}1.08$&	$\phantom{-}1.09\;(0.06)$&	0.06 & 0.95 &  &  &  \\ 
$\xi_0$ & $-1.58$ & $-1.58\;(0.07)$ & 0.06 & 0.95 & & & \\
$\xi_Y$ & $\phantom{-}1.85$ & $\phantom{-}1.86\;(0.08)$ & 0.08 & 0.95 & & &\\
$\eta_0$ & $-0.75$ & $-0.75\;(0.06)$ & 0.06 & 0.93 & & & \\
$\eta_{1}$ & $\phantom{-}0.15$&	$\phantom{-}0.15\;(0.03)$&	0.03& 0.95&  &  &  \\ 
$\beta_0$ & $\phantom{-}2.10$&	$\phantom{-}2.10\;(0.39)$&	0.40&	0.96&	$\phantom{-}2.12\;(0.35)$&	0.33 &	0.96 \\
$\beta_{\text{ODn}}$ & $-0.39$&	$-0.39\;(0.07)$& 0.07 & 0.94&	$-0.39\;(0.06)$&	0.06&	0.94 \\ 
$\beta_{\text{logVL}}$ & $-0.05$&	$-0.05\;(0.03)$ & 0.03&	0.95&	$-0.05\;(0.03)$&	0.03&	0.95 \\ 
$E(Y)$ & $\phantom{-}0.50$ & $\phantom{-}0.50$ & 0.02 &  & $\phantom{-}0.50$ & 0.02 & \\
\bottomrule
\end{tabular*}
\label{tab:sim2}
\end{table}

\begin{table}[!ht]
\caption{Average parameter estimate, average standard error, standard deviation, and coverage probability of 95\% confidence interval over 500 replicates. Training sample size = test sample size = contingency table sample size = 1000. We generate the contingency table for Sheng's method using single covariate—viral load (VL) or ODn.}
\scriptsize
\centering
\begin{tabular*}{\hsize}{@{}@{\extracolsep{\fill}}lcccccccccc@{}}
\toprule
& & \multicolumn{3}{c}{Our method} & \multicolumn{3}{c}{Sheng's method (VL)} & \multicolumn{3}{c}{Sheng's method (ODn)}\\
\cmidrule{3-5}
\cmidrule{6-8}
\cmidrule{9-11}
 & true value & estimate (se) & sd & coverage & estimate (se) & sd & coverage & estimate (se) & sd & coverage\\ 
\midrule
$\alpha$ & $\phantom{-}1.08$&	$\phantom{-}1.09\;(0.06)$&	0.06 & 0.95&  &  &  &  &  & \\ 
$\xi_0$ & $-1.58$&	$-1.58\;(0.07)$&	0.06 & 0.95&  &  &  &  &  & \\ 
$\xi_Y$ & $\phantom{-}1.85$&	$\phantom{-}1.86\;(0.08)$&	0.08 & 0.95&  &  &  &  &  & \\ 
$\eta_{0}$ & $-0.75$&	$-0.75\;(0.06)$&	0.06 & 0.93 &  &  & &  &  &  \\ 
$\eta_{1}$ & $\phantom{-}0.15$&	$\phantom{-}0.15\;(0.03)$&	0.03 & 0.95&  &  &  &  &  & \\ 
$\beta_0$ & $\phantom{-}2.10$&	$\phantom{-}2.10\;(0.39)$&	0.40&	0.96&	 $\phantom{-}0.99\;(0.30)$ & 0.30 & 0.04 & $\phantom{-}1.72\;(0.25)$&	0.23 &	0.67 \\
$\beta_{\text{ODn}}$ & $-0.39$&	$-0.39\;(0.07)$& 0.07 & 0.94&	&	&	 & $-0.42\;(0.06)$ & 0.05&	0.94\\ 
$\beta_{\text{logVL}}$ & $-0.05$&	$-0.05\;(0.03)$ & 0.03 & 0.95&	$-0.10\;(0.03)$&	0.03&	0.57 &  &  & \\ 
$E(Y)$ & $\phantom{-}0.50$ & $\phantom{-}0.50$ & 0.02 &  & $\phantom{-}0.50$ & 0.02 & &0.50  & 0.02 & \\
\bottomrule
\end{tabular*}
\label{tab:sim3}
\end{table}

\begin{table}[!ht]
\caption{Average parameter estimate, average standard error, standard deviation, and coverage probability of 95\% confidence interval over 500 replicates. Training sample size = test sample size = contingency table sample size = 1000.}
\scriptsize
\centering
\begin{tabular*}{\hsize}{@{}@{\extracolsep{\fill}}lcccccccccc@{}}
\toprule
& \multicolumn{2}{c}{True value} & \multicolumn{3}{c}{Our method} & \multicolumn{3}{c}{Sheng's method}\\
\cmidrule{2-3}
\cmidrule{4-6}
\cmidrule{7-9}
& primary data & external data & estimate (se) & sd & coverage & estimate (se) & sd & coverage \\ 
\midrule
$\alpha$ &  $\phantom{-}1.08$&& 	$\phantom{-}1.09\;(0.06)$&	0.06 & 0.95&  &     \\ 
$\xi_0$ & $-1.58$&& 	$-1.58\;(0.07)$&	0.06 & 0.95&  &    \\ 
$\xi_Y$ & $\phantom{-}1.85$&	& $\phantom{-}1.86\;(0.08)$&	0.08 & 0.95&  &    \\ 
$\eta_{0}$ & $-0.75$&& 	$-0.75\;(0.06)$&	0.06 & 0.93 &  &    \\ 
$\eta_{1}$ & $\phantom{-}0.15$&	& $\phantom{-}0.15\;(0.03)$&	0.03 & 0.95&  &   \\ 
$\beta_0$ & $\phantom{-}2.10$&	$\phantom{-}2.10$& $\phantom{-}2.10\;(0.39)$&	0.40&	0.96&	 $\phantom{-}2.11\;(0.44)$ & 0.44 & 0.95  \\
$\beta_{\text{ODn}}$ & $-0.39$&	$-0.52$& $-0.39\;(0.07)$& 0.07 & 0.94&	$-0.52\;(0.08)$ &	0.08 & 0.63	 \\ 
$\beta_{\text{logVL}}$ & $-0.05$& $\phantom{-}0.17$ & 	$-0.05\;(0.03)$ & 0.03 & 0.95&	$\phantom{-}0.17\;(0.04)$&	0.04 & 0\\ 
$E(Y)$ & $\phantom{-}0.50$ & $\phantom{-}0.88$ & $\phantom{-}0.50$ & 0.02 &  & $\phantom{-}0.81$ & 0.01  \\
\bottomrule
\end{tabular*}
\label{tab:sim4}
\end{table}

\begin{table}[!ht]
\caption{Average parameter estimate, average standard error, standard deviation, and coverage probability of 95\% confidence interval over 500 replicates. Training sample size of our method is 1000 while, on average, 452 of them have an observed response. Test sample size is 1000.}
\centering
\begin{tabular*}{\hsize}{@{}@{\extracolsep{\fill}}lccccccc@{}}
\toprule
& & \multicolumn{3}{c}{Our method} & \multicolumn{3}{c}{Naive logistic regression}\\
\cmidrule{3-5}
\cmidrule{6-8}
& true value & estimate (se) & sd & coverage & estimate (se) & sd & coverage \\ 
\midrule
$\alpha$     & $\phantom{-}1.07$  & $\phantom{-}1.09\;(0.06)$  & 0.06 & 0.93 &                             &      &      \\
$\xi$        & $-1.59$ & $-1.60\;(0.07)$  & 0.06 & 0.95 &                             &      &      \\
$\xi_Y$       & $\phantom{-}1.83$  & $\phantom{-}1.84\;(0.08)$  & 0.09 & 0.95 &                             &      &      \\
$\eta_0$      & $-0.74$ & $-0.82\;(0.06)$ & 0.06 & 0.78 &                             &      &      \\
$\eta_1$      & $\phantom{-}0.15 $ & $\phantom{-}0.12\;(0.02)$  & 0.02 & 0.78 &                             &      &      \\
$\beta_0$   & $\phantom{-}0.02$  & $\phantom{-}0.11\;(0.09)$  & 0.09 & 0.82 & $\phantom{-}0.62\;(0.10)$  & 0.11 & 0    \\
$\beta_\text{ODn}$ & $-0.50$  & $-0.48\;(0.08)$ & 0.08 & 0.94 & $-0.50\;(0.11)$ & 0.11 & 0.94 \\
$\beta_\text{age}$ & $-0.29$ & $-0.27\;(0.08)$ & 0.08 & 0.95 & $-0.29\;(0.10)$ & $0.10$  & 0.95 \\
$E(Y)$      & $\phantom{-}0.50$   &        $\phantom{-}0.53$                      & $0.02$ &      &                           $\phantom{-}0.64$  & $0.02$ & \\
\bottomrule
\end{tabular*}
\label{tab:sim5}
\end{table}

\begin{table}[!ht]
\caption{Average parameter estimate, average standard error, standard deviation, and coverage probability of 95\% confidence interval over 500 replicates. Training sample size of our method is 1000 while, on average, 452 of them have an observed response. Test sample size is 1000.}
\scriptsize
\centering
\begin{tabular*}{\hsize}{@{}@{\extracolsep{\fill}}lcccccccccc@{}}
\toprule
& & \multicolumn{3}{c}{Our method (forward telescoping)} & \multicolumn{3}{c}{Our method (backward telescoping)} & \multicolumn{3}{c}{Naive logistic regression}\\
\cmidrule{3-5}
\cmidrule{6-8}
\cmidrule{9-11}
 & true value & estimate (se) & sd & coverage & estimate (se) & sd & coverage & estimate (se) & sd & coverage\\ 
\midrule
$\alpha$ & $\phantom{-}1.07$&	$\phantom{-}1.11\;(0.06)$&	0.06 & 0.91& $\phantom{-}1.10\;(0.06)$ &  0.06&  0.93 & & & \\ 
$\xi_0$ & $-1.59$&	$-1.47\;(0.07)$&	0.07 & 0.56& $-1.56\;(0.07)$ & 0.07 & 0.93 &  &  & \\ 
$\xi_Y$ & $\phantom{-}1.83$&	$\phantom{-}1.78\;(0.08)$&	0.08 & 0.93& $\phantom{-}1.77\;(0.08)$ & 0.08 & 0.90 &  &  & \\ 
$\eta_{0}$ & $-0.74$&	$-0.78\;(0.06)$&	0.06 & 0.93 & $-0.73\;(0.06)$ & 0.05 & 0.95 &  &  &  \\ 
$\eta_{1}$ & $\phantom{-}0.15$&	$\phantom{-}0.16\;(0.03)$&	0.03 & 0.97&  $\phantom{-}0.16\;(0.03)$& 0.03 & 0.94 &  &  & \\ 
$\beta_0$ & $\phantom{-}0.02$&	$\phantom{-}0.08\;(0.09)$&	0.08&	0.88&	 $\phantom{-}0.03\;(0.09)$ & 0.08 & 0.96 & $\phantom{-}0.62\;(0.10)$&	0.10 &	0 \\
$\beta_{\text{ODn}}$ & $-0.50$&	$-0.49\;(0.08)$& 0.08 & 0.95 & $-0.51\;(0.08)$&	0.08&	0.95 & $-0.51\;(0.11)$ & 0.11 & 0.95\\ 
$\beta_{\text{age}}$ & $-0.29$&	$-0.29\;(0.08)$ & 0.08 & 0.95&	$-0.29\;(0.08)$ & 0.08 & 0.97 & $-0.30\;(0.11)$ & 0.10 & 0.95\\ 
$E(Y)$ & $\phantom{-}0.50$ & $\phantom{-}0.52$ & 0.02 &  & $\phantom{-}0.51$ & 0.02 & &0.64  & 0.02 & \\
\bottomrule
\end{tabular*}
\label{tab:sim6}
\end{table}

\begin{table}[!ht]
\caption{Average parameter estimate, average standard error, standard deviation, and coverage probability of 95\% confidence interval over 500 replicates. Training sample size of our method is 1000 while, on average, 469 of them have an observed response. Test sample size is 1000.}
\centering
\begin{tabular*}{\hsize}{@{}@{\extracolsep{\fill}}lccccccc@{}}
\toprule
& & \multicolumn{3}{c}{Our method} & \multicolumn{3}{c}{Naive logistic regression}\\
\cmidrule{3-5}
\cmidrule{6-8}
 & true value & estimate (se) & sd & coverage & estimate (se) & sd & coverage \\ 
\midrule
$\alpha$ & $\phantom{-}1.07$ & $\phantom{-}1.01\;(0.05)$ & 0.06 & 0.74 & & & \\
$\xi_0$ & $-1.59$ & $-1.64\;(0.07)$ & 0.07 & 0.87 & & & \\
$\xi_Y$ & $\phantom{-}1.83$ & $\phantom{-}1.71\;(0.09)$ & 0.11 & 0.72 & & & \\
$\eta_{0}$ & $-0.74$ &	$-0.61\;(0.05)$ &	$0.05$&	$0.29$ &   &  &  \\ 
$\eta_{1}$ & $\phantom{-}0.15$&	$\phantom{-}0.12\;(0.03)$ &	$0.03$&	$0.80$ &   &  &  \\
$\beta_0$ & $\phantom{-}0.02$&	$\phantom{-}0.08 \;(0.09)$ &	$0.09$ &	0.92&	$\phantom{-}0.66 \;(0.11)$&	0.11&	0 \\
$\beta_{\text{ODn}}$ & $-0.50$&	$-0.69 \;(0.09)$&	$0.10$ &	0.47&	$-0.99 \;(0.12)$&	0.12&	0.01 \\
$\beta_{\text{age}}$ & $-0.29$&	$-0.29 \;(0.08)$&	$0.08$ &	0.97&	$-0.29 \;(0.11)$&	0.11&	0.95 \\
$E(Y)$ & $\phantom{-}0.50$ & $\phantom{-}0.52$ & $0.02$ &  & $\phantom{-}0.63$ & 0.02 & \\
\bottomrule
\end{tabular*}
\label{tab:sim7}
\end{table}

\begin{figure}[!ht]
    \centerline{\includegraphics[width=\textwidth]{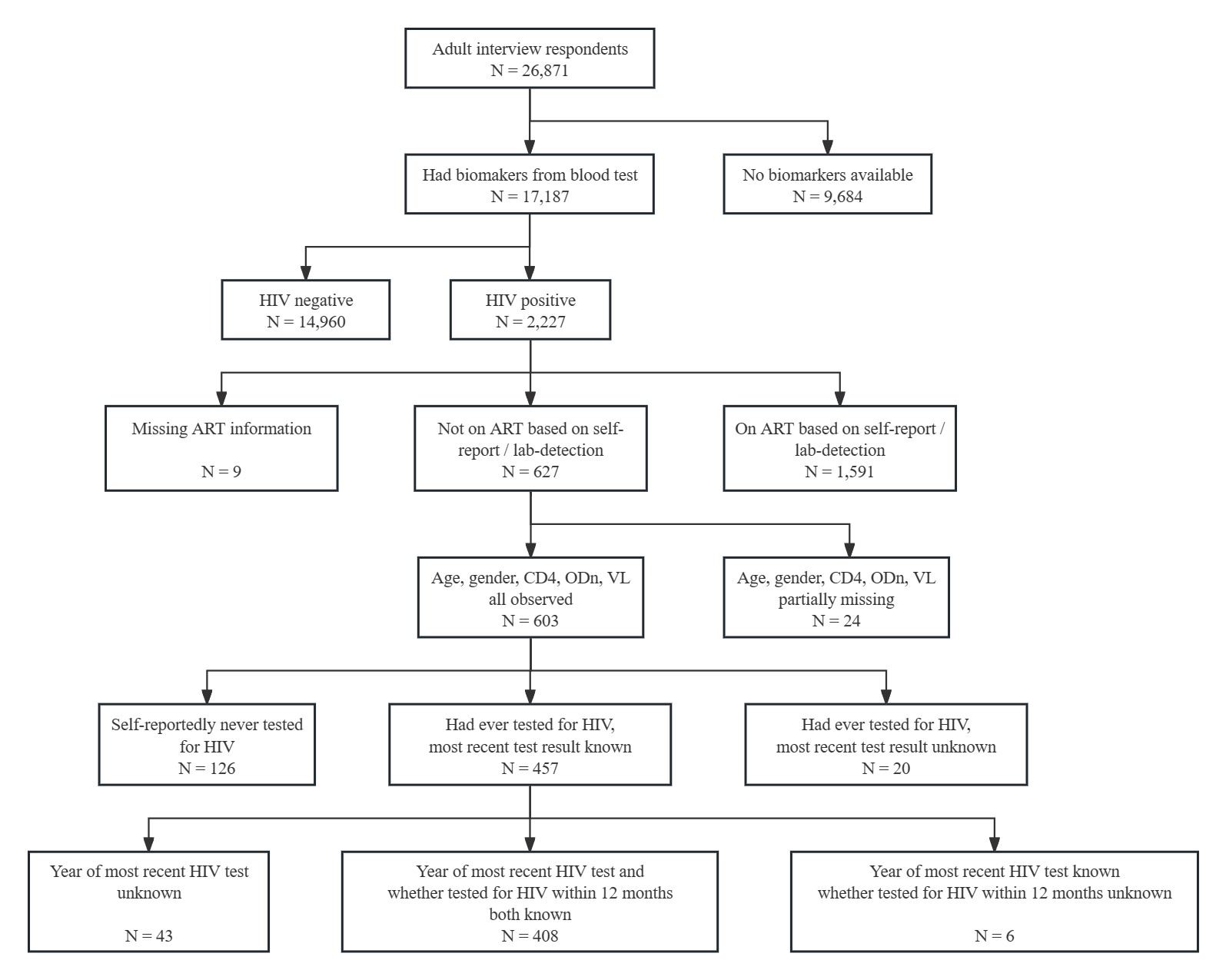}}
    \caption{CONSORT diagram describing sample selection.}
    \label{fig:consort}
\end{figure}

\begin{figure}[!ht]
    \centerline{\includegraphics[width=0.8\textwidth]{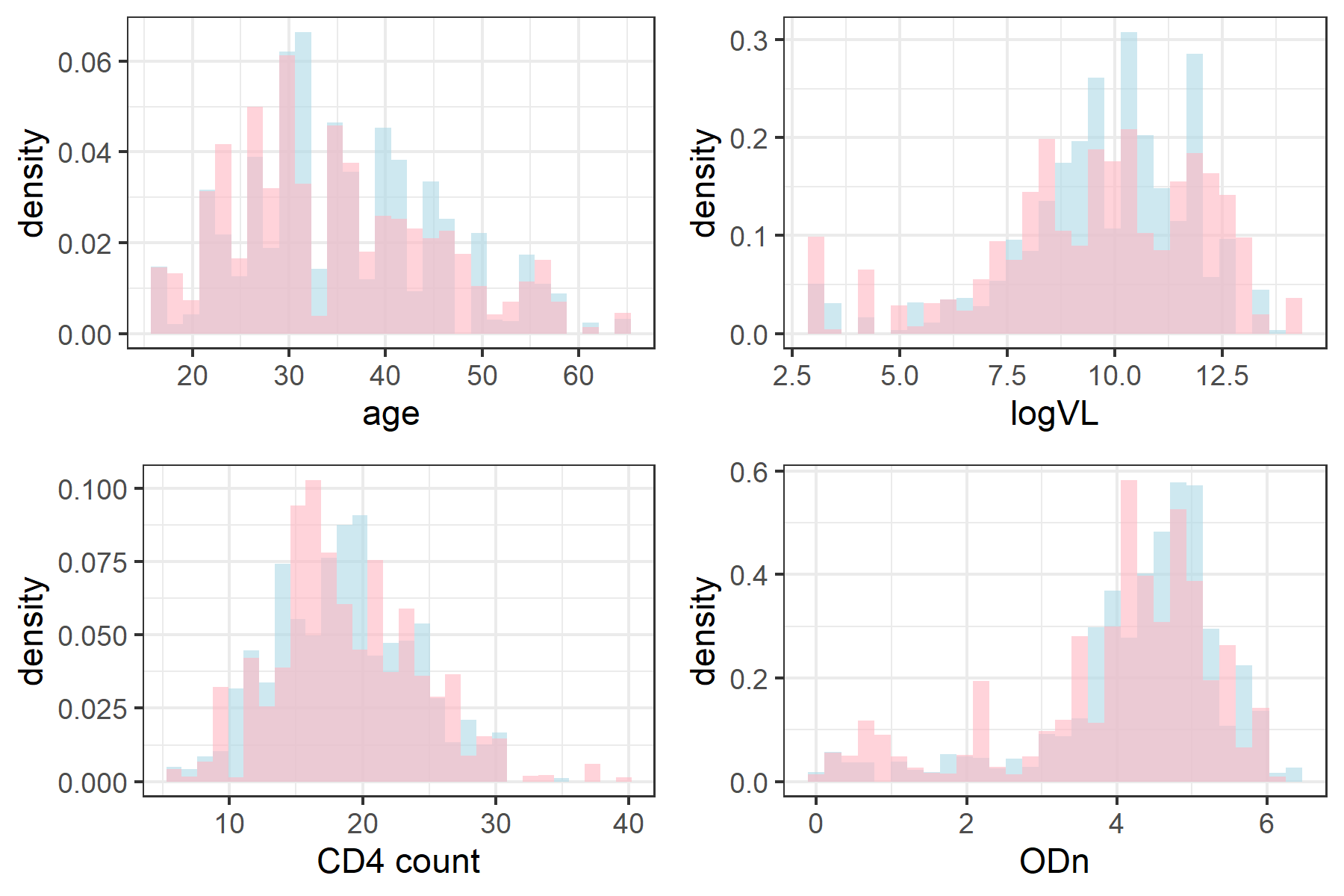}}
    \caption{Weighted histogram of some biomarkers grouped by whether $Y$ is observed, with light pink rectangles representing known response and light blue ones representing unknown response. People with known response tend to have smaller age and ODn, and have higher variation in logVL.}
    \label{fig:biasedsample}
\end{figure}

\begin{figure}
\centering
\begin{subfigure}{.5\textwidth}
  \centering
  \includegraphics[width=\linewidth]{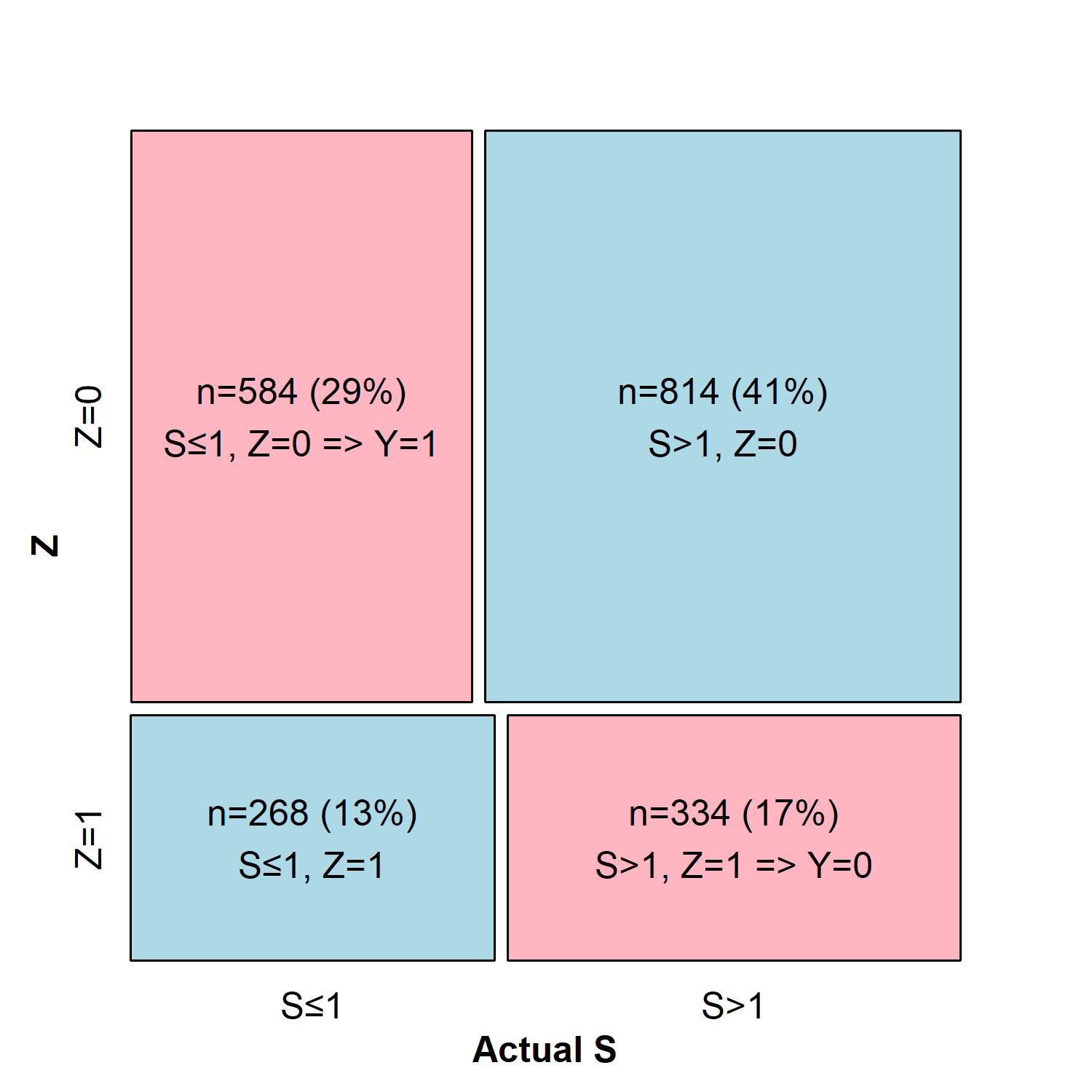}
  \caption{Actual value of $S$ and $Z$.}
  \label{fig:sub1}
\end{subfigure}%
\begin{subfigure}{.5\textwidth}
  \centering
  \includegraphics[width=\linewidth]{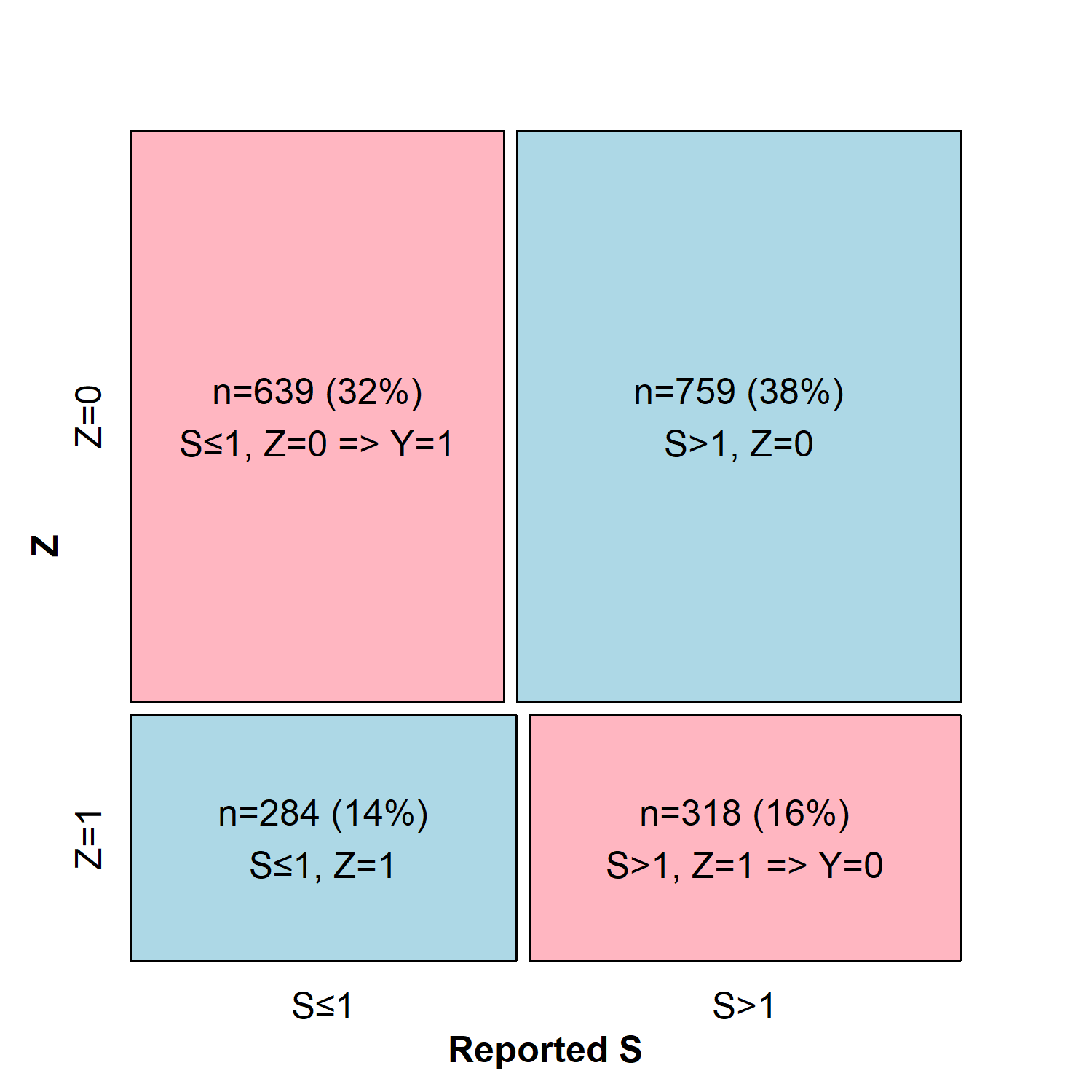}
  \caption{Reported $S$ is subject to forward telescoping.}
  \label{fig:sub2}
\end{subfigure}
\caption{Mosaic plot of whether the subject received a test within one year and whether the test result was positive in one replication of Simulation \Romannum{6}. Top left and bottom right blocks represent data with observed $Y$, while top right and bottom left ones represent unobserved $Y$.}
\label{fig:mosaicS}
\end{figure}

\FloatBarrier
\newpage
\bibliographystyle{biom}
\bibliography{biblio}